\documentclass[twocolumn,final,prb,showpacs,superscriptaddress,amsmath,amssymb,amsfonts,floatfix]{revtex4-1}
\usepackage{graphicx}
\usepackage{multirow}
\usepackage{epstopdf}
\usepackage{epsfig}
\usepackage{array}
\usepackage{verbatim}
\usepackage{latexsym}
\newcommand{\cucd}{CdCu$_2$(BO$_3$)$_2$}
\newcommand{\scbo}{SrCu$_2$(BO$_3$)$_2$}
\newcommand{\jd}{\ensuremath{J_{\text{d}}}}
\newcommand{\jti}{\ensuremath{J_{\text{t1}}}}
\newcommand{\jtii}{\ensuremath{J_{\text{t2}}}}
\newcommand{\jit}{\ensuremath{J_{\text{it}}}}
\newcommand{\jcd}{\ensuremath{J_{\text{Cd}}}}
\def\be{\begin{equation}}
\def\ee{\end{equation}}
\def\bea{\begin{eqnarray}}
\def\eea{\end{eqnarray}}

\def\vec{\mathbf}
\def\mc{\mathcal}

\DeclareMathAlphabet{\mathpzc}{OT1}{pzc}{m}{it}

\usepackage[colorlinks,breaklinks=true,bookmarks=false,citecolor=blue,linkcolor=red,urlcolor=blue]{hyperref}

\begin{document}

\title{Decorated Shastry-Sutherland lattice in the spin-$\frac12$ magnet
CdCu$_2$(BO$_3$)$_2$}

\author{O. Janson}
\email{janson@cpfs.mpg.de}
\affiliation{Max-Planck-Institut f\"{u}r Chemische Physik fester
Stoffe, D-01187 Dresden, Germany}

\author{I. Rousochatzakis}
\affiliation{Max-Planck-Institut f\"{u}r Physik komplexer Systeme, D-01187 Dresden, Germany}
\affiliation{Institute for Theoretical Solid State Physics, IFW Dresden, D-01171 Dresden, Germany}

\author{A.~A. Tsirlin}
\affiliation{Max-Planck-Institut f\"{u}r Chemische Physik fester
Stoffe, D-01187 Dresden, Germany}

\author{J. Richter}
\affiliation{Institut f\"{u}r Theoretische Physik, Universit\"{a}t
Magdeburg, D-39016 Magdeburg, Germany}

\author{Yu. Skourski}
\affiliation{Hochfeld-Magnetlabor Dresden, Helmholtz-Zentrum
Dresden-Rossendorf, D-01314 Dresden, Germany}

\author{H. Rosner}
\email{rosner@cpfs.mpg.de}
\affiliation{Max-Planck-Institut f\"{u}r Chemische Physik fester
Stoffe, D-01187 Dresden, Germany}

\date{\today}

\begin{abstract}
We report the microscopic magnetic model for the spin-$\frac12$
Heisenberg system CdCu$_2$(BO$_3$)$_2$, one of the few quantum magnets
showing the $\frac12$-magnetization plateau. Recent neutron diffraction
experiments on this compound $[$M. Hase~\emph{et~al}., Phys. Rev. B 80,
104405 (2009)$]$ evidenced long-range magnetic order, inconsistent
with the previously suggested phenomenological magnetic model of
isolated dimers and spin chains.  Based on extensive density-functional
theory band structure calculations, exact diagonalizations, quantum
Monte Carlo simulations, third-order perturbation theory, as well as high-field
magnetization measurements, we find that the magnetic properties of
CdCu$_2$(BO$_3$)$_2$ are accounted for by a frustrated quasi-2D magnetic
model featuring four inequivalent exchange couplings: the leading
antiferromagnetic coupling $J_{\text{d}}$ within the structural
Cu$_2$O$_6$ dimers, two interdimer couplings $J_{\text{t1}}$ and
$J_{\text{t2}}$, forming magnetic tetramers, and a ferromagnetic
coupling $J_{\text{it}}$ between the tetramers.  Based on comparison to
the experimental data, we evaluate the ratios of the leading couplings
$J_{\text{d}}$\,:\,$J_{\text{t1}}$\,:\,$J_{\text{t2}}$\,:\,$J_{\text{it}}$\,=\,1\,:\,0.20\,:\,0.45\,:\,$-$0.30,
with \jd\ of about 178\,K.  The inequivalence of $J_{\text{t1}}$ and
$J_{\text{t2}}$ largely lifts the frustration and triggers long-range
antiferromagnetic ordering.  The proposed model accounts  correctly for
the different magnetic moments localized on structurally inequivalent Cu
atoms in the ground-state magnetic configuration. We extensively analyze
the magnetic properties of this model, including a detailed description
of the magnetically ordered ground state and its evolution in magnetic
field with particular emphasis on the $\frac12$-magnetization plateau.
Our results establish remarkable analogies to the Shastry-Sutherland
model of SrCu$_2$(BO$_3$)$_2$, and characterize the closely related
CdCu$_2$(BO$_3$)$_2$ as a material realization for the spin-$\frac12$
decorated anisotropic Shastry-Sutherland lattice.
\end{abstract}


\maketitle

\section{Introduction}
\begin{figure}[tbp]
\includegraphics[width=8.6cm]{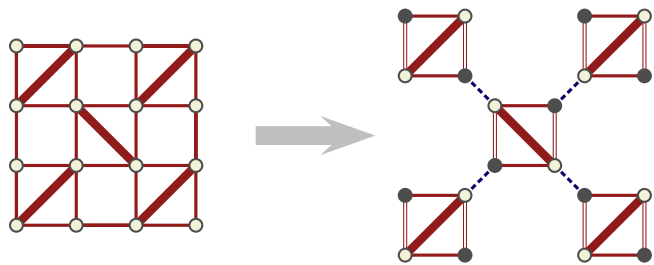}
\caption{\label{F-ShSu}(Color online) Left panel: the original
Shastry-Sutherland model with the dimer-like coupling $J$ (thick lines) and the
coupling $J'$ (thin lines) along the edges of the square lattice. Right panel:
the decorated anisotropic Shastry-Sutherland model in \cucd. The horizontal and
vertical couplings ($J'$ in the original model) are not equivalent, leading to
\jti\ (double line) and \jtii\ (single line). In addition, the coupling \jti\
(dashed line) tiles the lattice into four-spin units (tetramers). Filled
circles denote the decorating spins.} \end{figure}

Among the great variety of magnetic ground states (GSs), only few can be rigorously
characterized using analytical considerations. To these exceptional cases
belong, e.g., the Heisenberg zigzag chain model at the so-called
Majumdar--Ghosh point,\cite{FHC_AFAF_MG_I, *FHC_AFAF_GS_2} a class of Kitaev
models,\cite{Kitaev_model} as well as the dimerized phase of the Heisenberg
model on the Shastry-Sutherland lattice.\cite{Shastry_Suth} The latter
describes the magnetism of $S$\,=\,1/2 spins on a square lattice, with two
inequivalent exchange couplings: (i) a dimer-like coupling $J$ connecting the
spins pairwise along the diagonals of the square lattice, and (ii) the coupling
$J'$ along the edges of the square lattice (Fig.~\ref{F-ShSu}, left). For the
ratio of antiferromagnetic (AFM) couplings $J'$/$J$\,$<$\,0.7, the GS 
is an exact product of singlets residing on the $J$
bonds.\cite{SCBO_GS_ED}

The Shastry-Sutherland model was first studied
theoretically,\cite{Shastry_Suth} and only two decades later the spin
$S$\,=\,$\frac12$ magnet \scbo\ was claimed to be an experimental
realization of this model.\cite{SCBO} In the following years, numerous
experimental and theoretical studies disclosed a rather complex magnetic
behavior of this compound, including plateaux at 1/8, 1/4, and 1/3 of
the saturation magnetization (Ref.~\onlinecite{SCBO_MH}) and an
intricate coupling between magnetic and lattice degrees of
freedom.\cite{SCBO_lowT_str}  For the experimentally defined ratio of
the leading couplings $J'$/$J$\,=\,0.64
(Refs.~\onlinecite{SCBO_chiT_fit} and \onlinecite{SCBO_theory_review}),
the system is already in the dimerized singlet phase, but very close to
the quantum critical point.\cite{[{See, e.g.,
Ref.~\onlinecite{SCBO_theory_review}, }] [{}] SCBO_GS_SE, *SCBO_CCM} To
access the critical point, further tuning of the system is required. For
example, external pressure leads to at least one phase
transition at low temperatures,\cite{SCBO_NMR_pressure} as evidenced by nuclear magnetic
resonance (NMR) studies,\cite{[{For a recent review, see }] [{}]
SCBO_NMR_review} while a recent theoretical model\cite{SCBO_press_simul}
with two inequivalent dimer couplings $J$ predicts an unanticipated,
effectively one-dimensional Haldane
phase.\cite{SCBO_press_simul,SCBO_1D_model} 

An alternative way to tune the magnetic couplings is the directed substitution
of nonmagnetic atoms or anionic groups. The impact of this substitution on the
magnetic coupling regime can be very different. The simplest effect is a bare
change of the energy scale, which does not affect the physics, as for instance,
in the spin-$\frac12$ uniform-chain compounds Sr$_2$Cu(PO$_4$)$_2$ and
Ba$_2$Cu(PO$_4$)$_2$.\cite{HC_Sr2CuPO42_Ba2CuPO42_str_chiT_DTA,
*HC_Sr2CuPO42_DFT_chiT_fit} Another possibility is alteration of the
leading couplings, as in the frustrated square lattice
systems $AA'$VO(PO$_4$)$_2$ (Ref.~\onlinecite{FSL_anisotropic}) or the kagome
lattice compounds kapellasite Cu$_3$Zn(OH)$_6$Cl$_2$ and haydeeite
Cu$_3$Mg(OH)$_6$Cl$_2$.\cite{kapel_synth, *kapel_hayd_DFT,
*kapel_hayd_str_neutr} Such systems have excellent potential for
exploring magnetic phase diagrams. Finally, in some cases, chemical
substitutions drastically alter the underlying magnetic model, as shown
recently for the family of Cu$_2A_2$O$_7$ compounds
($A$\,=\,P,\,As,\,V).\cite{Cu2P2O7_DFT_chiT_MH, *Cu2V2O7_DFT, *Cu2As2O7}

However, the most common effect of chemical substitution is a structural
transformation, which apparently leads to a major qualitative change in the
magnetic couplings. For instance, the Se--Te substitution, accompanied by a
major structural reorganization, transforms the Heisenberg-chain system
CuSe$_2$O$_5$ (Ref.~\onlinecite{CuSe2O5}) into the spin dimer compound
CuTe$_2$O$_5$
(Ref.~\onlinecite{CuTe2O5_chiT_ESR_EHTB,*CuTe2O5_DFT_fchiT_fCpT_fMH}).
Likewise, the La--Bi substitution in the 2D square lattice system La$_2$CuO$_4$
(Ref.~\onlinecite{HTSC_Pickett}) leads to the three-dimensional magnet
Bi$_2$CuO$_4$ (Ref.~\onlinecite{Bi2CuO4_DFT}), etc.

Coming back to the $S$\,=\,$\frac12$ Shastry-Sutherland compound \scbo,
the directed substitution of nonmagnetic structural elements is an
appealing tool to explore the phase space of the Shastry-Sutherland
model.\cite{SCBO_Ba_doping,
*SCBO_met_doping_chiT,*SCBO_Zn_Si_doping_chiT_MH, *SCBO_Mg_doping_INS}
One possible way is the replacement of Sr$^{2+}$ by another divalent
cation. Indeed, the compound \cucd\ exists, and its magnetic properties
have been investigated experimentally.\cite{CC_Cu2CdB2O6_chiT_CpT_MH,
CC_Cu2CdB2O6_ESR, CC_Cu2CdB2O6_NPD_MH} In contrast to \scbo, the
magnetic GS of \cucd\ is long-range ordered, while the magnetization
$M(H)$ curve exhibits a plateau at one-half of the saturation
magnetization.\cite{CC_Cu2CdB2O6_chiT_CpT_MH}

The substitution of Sr by Cd in \scbo\ is accompanied by a structural
transformation.  Thus, the initial tetragonal symmetry is reduced to
monoclinic, engendering two independent magnetic sites: Cu(1) and
Cu(2).  The presence of Cu(1) residing in the structural dimers, similar
to \scbo, and the relatively short Cu(2)--Cu(2) connections along $c$ motivated the authors of
Ref.~\onlinecite{CC_Cu2CdB2O6_chiT_CpT_MH}  to advance a tentative
magnetic model comprising isolated magnetic dimers of Cu(1) pairs and
infinite chains built by Cu(2) spins. This model was in agreement with
the experimental magnetic susceptibility $\chi(T)$, magnetization
$M(H)$, and electron spin resonance (ESR) data.  However, later neutron
diffraction (ND) studies challenged this model, since the ordered
magnetic moments localized on Cu(1) amount to 0.45\,$\mu_{\text{B}}$
compared to 0.83\,$\mu_{\text{B}}$ on Cu(2).\cite{CC_Cu2CdB2O6_NPD_MH}
Therefore, the Cu(1) dimers do not form a singlet state with zero
ordered moment, but rather experience a sizable coupling to Cu(2). This
coupling is ferromagnetic, as inferred from the magnetic structure, and
contrasts with the AFM coupling assumed in the previous
studies.\cite{CC_Cu2CdB2O6_chiT_CpT_MH} 

The tentative magnetic model from Ref.~\onlinecite{CC_Cu2CdB2O6_chiT_CpT_MH}
was founded on the assumption that the strongest magnetic couplings are
associated with the shortest Cu--Cu interatomic distances. However, such 
an approach fully neglects the mutual orientation of magnetic orbitals, which
play a decisive role for the magnetic coupling regime. Moreover, the
experimental results suggest that the actual spin model of \cucd\ is more
complex than the simple ``chain + dimer'' scenario.

In this paper, we investigate the magnetic properties of \cucd\ on a
microscopic level, using density-functional-theory (DFT) band structure
calculations, exact diagonalization (ED) and quantum Monte Carlo (QMC)
numerical simulations, as well as analytical low-energy perturbative expansions
and linear spin-wave theory.  We show that the magnetism of \cucd\ can be
consistently described by a two-dimensional (2D) frustrated isotropic model
(crystalline and exchange anisotropy effects are neglected) with four
inequivalent exchange couplings, topologically equivalent to the decorated
anisotropic Shastry-Sutherland lattice (see Fig.~\ref{F-ShSu}, right).  Therefore,
the chemical similarity between \scbo\ and \cucd\  is retained on the
microscopic level.  The dominance of the exchange interaction \jd\ within the
structural Cu(1) dimers allows to describe the magnetic properties within an
effective low-energy model for the Cu(2) sites only. This effective model
explains the sizable staggered magnetization of Cu(1) dimers despite the large
\jd, and unravels the nature of the wide 1/2-magnetization plateau in \cucd.
The experimental data, both original and taken from the literature, are
reconsidered in terms of the DFT-based microscopic model and the effective
model. As a result, we find an excellent agreement between the simulated and
measured quantities. 

This paper is organized as follows. Sec.~\ref{S-str} is devoted to the
discussion of the structural peculiarities of \cucd. The methods used in
this study are presented in Sec.~\ref{S-method}. Then, the microscopic
magnetic model is evaluated using DFT band structure calculations, and
its parameters are refined using ED fits to the experiments
(Sec.~\ref{S-model-eval}). The magnetic properties of \cucd\ are in the
focus of Sec.~\ref{S-model-prop}.  We show that the magnetism of \cucd\
is captured by  an effective low-energy description of the Cu(2) spins
(on the Shastry-Sutherland lattice), which are further studied using
linear spin-wave theory as well as QMC, and corroborated by ED for the microscopic
magnetic model. Finally, in Sec.~\ref{S-sum} a summary and a short
outlook are given.

\section{\label{S-str}Crystal structure}
\begin{figure}[tbp]
\includegraphics[width=8.6cm]{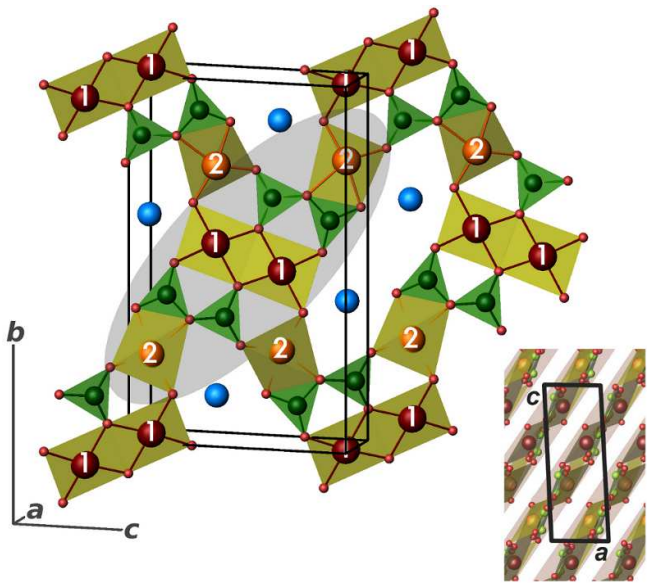}
\caption{\label{F-str}(Color online) The crystal structure of \cucd\
consists of structural Cu(1)$_2$O$_6$ dimers, Cu(2)O$_4$ distorted
plaquettes, BO$_3$ triangles connected pairwise as well as isolated Cd
atoms (spheres). Oxygen atoms are shown as small spheres located at the
vertices of Cu(1)O$_4$, Cu(2)O$_4$ and BO$_3$ polyhedra. The numbers
``1'' and ``2'' denote Cu(1) and Cu(2), respectively. A \emph{magnetic}
tetramer (see Sec.~\ref{S-DFT}) is shaped by a gray oval. Right bottom:
position of the magnetic layers (shaded) in the crystal structure of
\cucd.}
\end{figure}

The low-temperature crystal structure of \cucd\ is reliably established based
on synchrotron x-ray and neutron diffraction data.\cite{CC_Cu2CdB2O6_NPD_MH} The basic structural
elements of \cucd\ are magnetic layers (Fig.~\ref{F-str}) that stretch almost
parallel to $(\bar{1}02)$ planes (Fig.~\ref{F-str}, right bottom). These layers
are formed by Cu--O polyhedra coupled by B$_2$O$_5$ units.

Although Cd$^{2+}$ is isovalent to Sr$^{2+}$, the substantial difference
in their ionic radii (1.26\,\r{A} and 0.95\,\r{A},
respectively)\cite{ionic_radii} gives rise to different structural
motives in the two systems.  Thus, in sharp contrast with \scbo, the
monoclinic crystal structure of \cucd\ comprises two inequivalent
positions for Cu atoms: Cu(1) and Cu(2). The local environment of these
two sites is different: Cu(1) forms Cu(1)O$_4$ plaquettes that in turn
build structural Cu(1)$_2$O$_6$ dimers, resembling SrCu$_2$(BO$_3$)$_2$,
while Cu(2) atoms form tetrahedrally distorted Cu(2)O$_4$ plaquettes
that share a common O atom with the Cu(1)$_2$O$_6$ dimers
(Fig.~\ref{F-str}). As a result, structural Cu(1)$_2$Cu(2)$_2$O$_{12}$
tetramers are formed.  Interestingly, the structural tetramers do not
coincide with the magnetic tetramers (shaded oval in Fig.~\ref{F-str})
that are the basic units of the spin lattice (Sec.~\ref{S-DFT}).

Large cavities between the neighboring magnetic tetramers accommodate Cd
atoms.  As a result, the space between the magnetic layers is rather
narrow, leading to short interatomic distances ($\sim$3.4\,\r{A})
between the Cu atoms belonging to neighboring magnetic layers. On the
basis of such short Cu--Cu distance, the relevance of the respective
interlayer coupling (the chain-like coupling $J_3$ in the notation of
Ref.~\onlinecite{CC_Cu2CdB2O6_chiT_CpT_MH}) has been
conjectured.\cite{CC_Cu2CdB2O6_chiT_CpT_MH} However, the orientation of
magnetic Cu(1)$_2$O$_6$ and Cu(2)O$_4$ polyhedra (almost coplanar to the
magnetic layers) excludes an appreciable interlayer coupling.  This
quasi-2D scenario is readily confirmed by our DFT calculations
(Sec.~\ref{S-DFT}).

The evaluation of the intralayer coupling based on structural
considerations is more involved. In many cuprates, the leading exchange
couplings are typically associated with the Cu--O$_{\text{plaq}}$--Cu
connections, i.e., when the neighbouring plaquettes share a common
corner (one O$_{\text{plaq}}$ atom) or a common edge (two O$_{\text{plaq}}$
atoms).\footnote{It is especially important that the respective O atom
belongs to a magnetic plaquette. Otherwise, the magnetic exchange would be
very small. For instance, an octahedral CuO$_6$ coordination is
typically distorted with a formation of four short Cu--O$_{\text{plaq}}$
bonds (a plaquette) and two longer Cu--O$_{\text{apic}}$ (apical) bonds.
The latter bonds are perpendicular to the magnetically active Cu
$3d_{x^-y^2}$ orbital, thus the superexchange along
Cu--O$_{\text{apic}}$--Cu paths can be very small.} In the crystal
structure of \cucd, both types of connections are present. 

To estimate the respective magnetic exchange couplings, the
Goodenough--Kanamori rules\cite{GKA_1, *GKA_2} are typically applied.
Using a closely related approach, Braden \emph{et~al.}\cite{J_CuOCu}
evaluated the dependence of the magnetic exchange $J$ on the Cu--O--Cu
angle. According to their results, for the Cu--O--Cu angle of
90$^{\circ}$, the resulting total coupling $J$, being a sum of AFM
$J^{\text{AFM}}$ (positive) and FM $J^{\text{FM}}$ (negative)
contributions, is FM. An increase in this angle (i) decreases the
$J^{\text{FM}}$ contribution caused by Hund's coupling on the O
sites,\cite{FM_exchange_WF} and (ii) enhances the Cu--O--Cu
superexchange which in turn increases $J^{\text{AFM}}$. As a result, for
the Cu--O--Cu angle close to 97$^{\circ}$, $J^{\text{FM}}$ and
$J^{\text{AFM}}$ are balanced, thereby cancelling the total magnetic
exchange ($J$\,=\,0). Further increase in the Cu--O--Cu angle breaks
this subtle balance and gives rise to an overall AFM exchange, which
grows monotonically up to Cu--O--Cu\,=\,180$^{\circ}$.

Here, we try to apply this phenomenological approach and estimate the
respective exchange integrals. The intradimer Cu(1)--O--Cu(1) angle in
\cucd\ is 98.2$^{\circ}$, only slightly exceeding 97$^{\circ}$.
Therefore, a weak AFM exchange coupling could be expected. For the
Cu(1)--O--Cu(2) corner-sharing connections, the angle amounts to
117.5$^{\circ}$, which should give rise to a sizable AFM exchange.
However, our DFT+$U$ calculations deliver a magnetic model, which is in
sharp contrast with these expectations: the AFM coupling within the
structural dimers is sizable, while the coupling along the
Cu(1)--O--Cu(2) connections is much smaller and FM. The substantial
discrepancy between the simplified picture, based on structural
considerations only, and the microscopic model, evidences the crucial
importance of a microscopic insight for structurally intricate magnets
like \cucd.

\section{\label{S-method}Methods}
DFT band structure calculations were performed using the full-potential
code \textsc{fplo-8.50-32}.\cite{FPLO} For the structural input, we used
the atomic positions based on the neutron diffraction data at
1.5\,K: space group $P2_1/c$\,(14),
$a$\,=\,3.4047\,\r{A}, $b$\,=\,15.1376\,\r{A}, $c$\,=\,9.2958\,\r{A},
$\beta$\,=\,92.807$^{\circ}$.\cite{CC_Cu2CdB2O6_NPD_MH}  For the scalar-relativistic
calculations, the local-density approximation (LDA) parameterization of
Perdew and Wang has been chosen.\cite{PW92} We also cross-checked our
results by using the parameterization of Perdew, Burke and Ernzerhof
based on the generalized gradient approximation (GGA).\cite{PBE96} For
the nonmagnetic calculations, a $k$mesh of
30$\times$6$\times$12\,=\,2160 points (728 points in the irreducible
wedge) has been adopted. Wannier functions (WF) for the Cu
$3d_{x^2-y^2}$ states were evaluated using the procedure described in
Ref.~\onlinecite{FPLO_WF}. For the supercell magnetic DFT+$U$
calculations, two types of supercells were used: a supercell metrically
equivalent to the unit cell (sp.\ gr.\ $P1$, 4$\times$2$\times$2
$k$mesh) and a supercell doubled along $c$ (sp.\ gr.\ $P1$,
4$\times$1$\times$1 $k$mesh). Depending on the double-counting
correction (DCC),\cite{LDA_U_AMF_FLL,*LSDA+U_Pickett} the
around-mean-field (AMF) or the fully localized limit (FLL), we varied
the on-site Coulomb repulsion parameter $U_{3d}$ in a wide range:
$U_{3d}$\,=\,5.5--7.5\,eV  for the AMF and $U_{3d}$\,=\,8.5--10.5\,eV
for the FLL calculations,\footnote{Note that different $U_{3d}$ ranges
should be used for AMF and FLL, as explained in
Refs.~\onlinecite{Cu2P2O7_DFT_chiT_MH,*Cu2V2O7_DFT} and
\onlinecite{HC_dio_DFT_QMC_chiT}.} keeping $J_{3d}$\,=\,1\,eV.  The
local spin-density approximation (LSDA)+$U$ results were cross-checked
by GGA+$U$ calculations. All results were accurately checked for
convergence with respect to the $k$mesh. 

Full diagonalizations of the resulting Heisenberg Hamiltonian were
performed using the software package \textsc{alps-1.3}
(Ref.~\onlinecite{ALPS}) on the $N$\,=\,16 finite lattice with
periodic boundary conditions. Translational symmetries have been used
(the code does not allow for rotational symmetries).  Spin correlations
in the GS as well as the lowest-lying excitations were computed by
Lanczos diagonalization of $N$\,=\,32 sites finite lattices using the
code \textsc{spinpack}.\cite{spinpack} Translational 
symmetries and the commutation relation $[H,S_z]$\,=\,0 were used. The
finite lattices are depicted in Fig.~\ref{F-sisj}.  The QMC simulations
of the effective model were performed on finite lattices with up to
$N$\,=\,2304 spins with periodic boundary conditions using the
\texttt{looper}\cite{loop} algorithm from the \texttt{ALPS}
package.\cite{ALPS} For $T$\,=\,0.05\,$J$\,$\simeq$\,0.0031\,\jd
($\simeq$\,0.55\,K), we used 40\,000 loops for thermalization and
400\,000 loops after thermalization. 

A powder sample of \cucd\ was prepared by annealing a stoichiometric
mixture of CuO, CdO, and B$_2$O$_3$ in air at 750~$^{\circ}$C for 96
hours. The sample contained trace amounts of unreacted CuO and
Cd$_2$B$_2$O$_5$ (below 1~wt.\% according to the Rietveld refinement),
as evidenced by powder x-ray diffraction (Huber G670 camera,
CuK$_{\alpha1}$ radiation, ImagePlate detector,
$2\theta$\,=\,3--100$^{\circ}$ angle range). To improve the sample
quality, we performed additional annealings that, however, resulted in a
partial decomposition of the target \cucd\ phase. Both impurity phases
reveal a low and nearly temperature-independent magnetization within the
temperature range under investigation. Therefore, they do not affect any
of the results presented below.

The magnetic susceptibility ($\chi$) was measured with an MPMS SQUID
magnetometer in the temperature range $2-380$~K in applied fields up to
5~T. The high-field magnetization curve was measured in pulsed magnetic
fields up to 60\,T at a constant temperature of 1.5\,K using the
magnetometer installed at the Dresden High Magnetic Field Laboratory.
Details of the experimental procedure can be found in
Ref.~\onlinecite{tsirlin2009}.

\section{\label{S-model-eval}Evaluation of a microscopic magnetic model}
\subsection{\label{S-DFT}Band structure calculations} 
\begin{figure}[htbp]
\includegraphics[width=8.6cm]{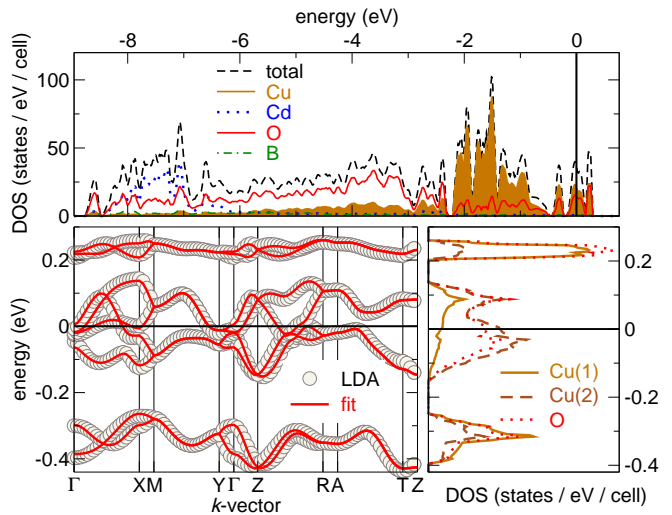}
\caption{\label{F-dos}(Color online) Top: LDA valence band of \cucd.
Atom-resolved DOS (number of states per eV and crystallographic unit cell) are
shown. Bottom left: LDA band structure of the well-separated eight-band complex
at the Fermi level and the WF tight-binding fit within an effective one-orbital
model. Bottom right: atom-resolved DOS of the eight-band complex.}
\end{figure}

We begin our
analysis with the evaluation of the magnetically relevant orbitals and
couplings. The LDA yields a valence band with a width of about 9\,eV,
dominated by O $2p$ states (Fig.~\ref{F-dos}, top panel).
Interestingly, the lower part of the valence band exhibits sizable
contribution from the Cd $4d$ states, hinting at an appreciable
covalency of Cd--O bonds. The energy range between $-3$ and $0.5$\,eV is
dominated by Cu $3d$ states. The presence of bands crossing the Fermi
level evidences a metallic GS, which contrasts with the emerald-green
color of \cucd. This well-known shortcoming of the LDA and the GGA
originates from underestimation of strong correlations, intrinsic for
the $3d^9$ electronic configuration of Cu$^{2+}$. To remedy this
drawback, we use two approaches accounting for the missing part of
correlations: (i) on the model level, via mapping the LDA bands onto a
Hubbard model (model approach); and (ii) directly within a DFT code, by
adding an energy penalty $U_{3d}$ for two electrons occupying the same
orbital (DFT+$U$ approach).

\begin{figure}[!tb]
\includegraphics[width=8.6cm]{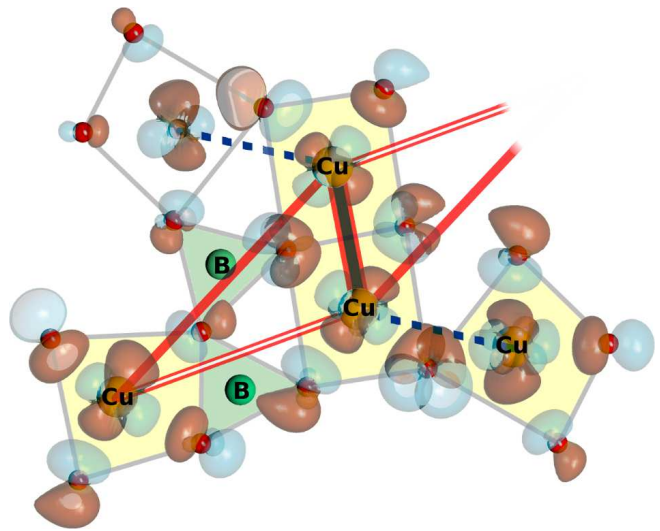}
\caption{\label{F-wf}(Color online) Wannier functions for the Cu
$3d_{x^2-y^2}$ states. Light and dark areas (colors) denote different
phases of the functions.  Relevant exchange couplings are depicted by
lines: thick (red-black) lines (\jd), thinner solid (\jtii) and double
(\jti) lines, as well as dashed lines (\jit). Note the Cu--O--O--Cu
superexchange paths via the BO$_3$ triangles.}
\end{figure}

The states relevant for magnetism are located in a close vicinity of the
Fermi level. Common for cuprates, these states form a band complex,
well-separated from the rest of the valence band. This band complex,
depicted in Fig.~\ref{F-dos} (bottom panels), comprises eight bands, in
accord with the eight Cu atoms in a unit cell. The projection of these
states onto a local coordinate system\footnote{In this coordinate
system, the $x$-axis runs from Cu to one of the O atoms, while the
$z$-axis is perpendicular to the plaquette plane.} readily yields their
orbital character: typical for cuprates, these states correspond to the
antibonding $\sigma$-combination of Cu~$3d_{x^2-y^2}$ and O~$2p$
orbitals. To visualize this combination in real space, the WFs for Cu
$3d_{x^2-y^2}$ states can be used (Fig.~\ref{F-wf}). 

To evaluate the relevant couplings, we calculate hopping matrix elements of the
WFs and introduce them as parameters ($t_{i}$) into an effective one-orbital
tight-binding model. This parameterization is justified by an excellent fit
(Fig.~\ref{F-dos}, bottom left) to the LDA band dispersions. The relevant
couplings ($t_{i}$\,$>$\,20\,meV) are presented in Table~\ref{T-tJ} (third
column).

The transfer integrals $t_{i}$ can be subdivided into four groups,
according to their strength. The first group contains only one term
$t_{\text{d}}$ (``d'' stands for ``dimer''), which clearly dominates
over all other couplings, despite the respective Cu(1)--O--Cu(1) angle
being rather close to 90$^{\circ}$. The second group comprises two
terms, $t_{\text{t1}}$ and $t_{\text{t2}}$ (``t'' stands for
``tetramer'') that couple Cu(1) and Cu(2) atoms via Cu(1)--O--O--Cu(2)
paths (Fig.~\ref{F-wf}) within the magnetic tetramers. Although the
Cu(1)--Cu(2) distance for $t_{\text{t1}}$ is shorter, the respective
coupling is slightly smaller than $t_{\text{t2}}$. This difference plays
an important role for the magnetism of \cucd, as will be shown later.

The couplings $t_{\text{d}}$, $t_{\text{t1}}$, and $t_{\text{t2}}$ form
magnetic tetramers (shaded oval in Fig.~\ref{F-str}). These tetramers
are coupled to each other by two interactions forming the third group:
$t_{\text{it}}$ and $t_{\text{Cd}}$. The coupling $t_{\text{it}}$ (`it''
stands for ``intertetramer'') corresponds to the magnetic exchange $J_2$
between dimers and chains in the notation of
Ref.~\onlinecite{CC_Cu2CdB2O6_chiT_CpT_MH}. It is short-ranged, and runs
via the corner-sharing Cu(1)--O--Cu(2) connection. The second
intertetramer term $t_{\text{Cd}}$ couples Cu(1) and Cu(2) via Cd atoms
(not shown). Finally, the fourth group of transfer integrals includes
all other terms that are smaller than 20\,meV. Since these latter terms
play a minor role for the magnetic properties of \cucd, they are
disregarded in the following discussion.

\begin{table}[tb]
\caption{\label{T-tJ}
Cu--Cu distances ($d$, in~\r{A}), transfer integrals ($t_i$, in~meV),
antiferromagnetic $J_i^{\text{AFM}}$ (in~K) contributions, as well as
the total exchange integrals $J_i$ (in~K) for the leading couplings in
\cucd. The values of $J_i$ are evaluated within the DFT+$U$ method for
two different double-counting correction schemes with
$U_{3d}$\,=\,6.5\,eV for AMF and 9.5\,eV  for FLL.}
\begin{ruledtabular}
\begin{tabular}{c c c p{.01\textwidth} c r r p{.01\textwidth} r}
\multirow{2}{*}{exchange} & \multirow{2}{*}{$d$} & \multirow{2}{*}{$t_i$} &
& \multirow{2}{*}{$J_i^{\text{AFM}}$}  & & \multicolumn{3}{c}{$J_i$ (LSDA+$U$)} \\
& & & & & & AMF & & FLL \\ \hline
\jd   & 2.957 & 203 & & 425 & &  146 & &  202  \\
\jti  & 5.268 & 73  & &  55 & &   30 & &   29  \\
\jtii & 6.436 & 85  & &  75 & &   64 & &   55  \\
\jit  & 3.228 & 47  & &  23 & &$-$45 & & $-$135  \\
\jcd  & 6.450 & 41  & &  17 & &    5 & &    6   \\ \hline
\multicolumn{2}{c}{\jti\,:\,\jd~(\%)} & & & & &   20 & & 14  \\
\multicolumn{2}{c}{\jit\,:\,\jd~(\%)} & & & & & $-31$ & & $-67$ \\ 
\multicolumn{2}{c}{\jti\,:\,\jtii~(\%)} & & & & & 47 & & 53 \\ 
\end{tabular}
\end{ruledtabular}
\end{table}

To account for electronic correlations, neglected in the tight-biding
model, the transfer integrals $t_i$ are mapped onto a Hubbard model.
This way, under the conditions of half-filling and strong correlations
($U_{\text{eff}}$\,$\gg$\,$t_i$), the AFM contribution
$J_i^{\text{AFM}}$ to the exchange integrals can be estimated in
second-order perturbation theory:\cite{superexchange}
$J_i^{\text{AFM}}$\,=\,$4t_i^2/U_{\text{eff}}$.\footnote{Restriction to
the antiferromagnetic contribution arises from the one-orbital nature of
the tight-binding model, which can not account for the ferromagnetic
exchange due to the Pauli exclusion principle.} Here, $U_{\text{eff}}$
is an effective term which takes into account the correlations in the
magnetic Cu $3d_{x^2-y^2}$ orbitals, screened by the O $2p$ orbitals
(their strong hybridization can be seen in Fig.~\ref{F-wf}).  Although
the experimental evaluation of $U_{\text{eff}}$ is challenging (see,
e.g., Ref.~\onlinecite{HC_Sr2CuO3_XAS}), there is an empirical evidence
that $U_{\text{eff}}$ values in the 4--5\,eV range yield reasonable
results for Cu$^{2+}$
compounds.\cite{HC_Sr2CuPO42_DFT_chiT_fit,CuSe2O5,Bi2CuO4_DFT} Here, we
adopt $U_{\text{eff}}$\,=\,4.5\,eV to estimate $J_i^{\text{AFM}}$
(Table~\ref{T-tJ}, fourth column).

Resorting from $t_i$ to $J_i^{\text{AFM}}$ basically preserves the
picture of coupled tetramers with a strong dimer-like coupling.
However, the interatomic Cu-Cu distances for \jd\ and \jit\ are rather
small, thus sizable FM contributions to these couplings can be expected.
To account for the FM exchange, we perform DFT+$U$ calculations for
different magnetic supercells,\cite{[{Strong correlations inherent to
the Cu $3d^9$ electronic configuration ensure the applicability of the
DFT+$U$ methods, see, e.g.}] [{}] LDA+U_GW_compar}
and subsequently map the resulting total
energies onto a Heisenberg model.  This way, the FM contribution can be
evaluated by subtracting $J_i^{\text{AFM}}$ estimates from the values of
the total exchange $J_i$, obtained from DFT+$U$ calculations.  The
results of DFT+$U$ calculations typically depend on (i) the functional
(LSDA or GGA), (ii) the DCC scheme (AMF or FLL), and (iii) the value of
$U_{3d}$ used. \footnote{The relevant ranges of $U_{3d}$ were chosen
based on previous DFT+$U$ studies for related cuprate systems (e.g.,
Refs.~\onlinecite{CuSe2O5} and \onlinecite{Bi2CuO4_DFT}). The parameter
$U_{3d}$ describes the correlations in the $3d$ shell, only, thus its
value is larger than the value of $U_{\text{eff}}$.} Representative
values for the total exchange integrals $J_i$ are summarized in
Table~\ref{T-tJ}.

We first discuss common trends that do not depend on a particular
DFT+$U$ calculational scheme. First, despite the substantial FM
contribution, \jd\ preserves its role of the leading coupling. Second,
the FM contributions further enhance the difference between \jti\ and
\jtii.  This difference is crucial, since these two couplings are
frustrated, and the degree of frustration largely depends on their
ratio. Third, the interdimer coupling \jit\ is clearly FM, while a
sizable FM contribution to \jcd\ practically switches this coupling off. 

The DFT+$U$ results exhibit a rather strong dependence on the
calculational scheme and the parameters used. A comparative analysis of
the exchange values shows that the couplings \jd, \jti, \jtii, and \jcd\
yielded by GGA+$U$ are reduced by 20--30\,\% compared to the LSDA+$U$
estimates (for the same DCC and $U_{3d}$). The difference between the
LSDA+$U$ and GGA+$U$ results for \jit\ is slightly larger. However, the
ratios of the leading couplings, which determine the nature of the GS,
are very similar for both functionals.

The DCC scheme has a much larger impact on the values of exchange integrals.
In particular, FLL yields substantially larger \jit\ values than AMF.
In addition, the \jti/\jd\ and \jtii/\jd\ ratios are slightly reduced in FLL
compared to the AMF estimates (Table~\ref{T-tJ}).

The dependence of the DFT+$U$ results on $U_{3d}$ shows a clear and
robust trend: larger $U_{3d}$ lead to smaller absolute values of the
leading exchange couplings. However, the slope of the $J_i(U_{3d})$
dependence (not shown) slightly differs for different couplings. As a
result, the ratios of the couplings also depend moderately on the
$U_{3d}$ value.  

\begin{figure}[tb]
\includegraphics[width=8.6cm]{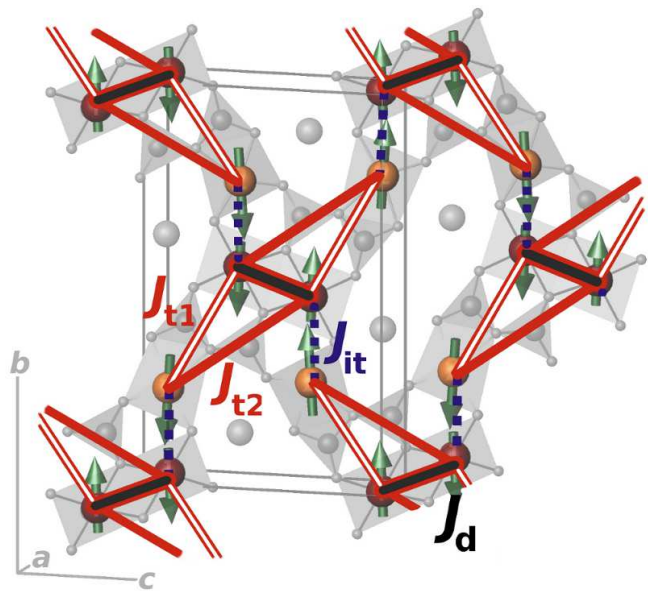}
\caption{\label{F-model}(Color online) Microscopic magnetic model for
\cucd. The leading magnetic couplings are denoted by lines. The
experimental magnetic structure (Ref.~\onlinecite{CC_Cu2CdB2O6_NPD_MH})
is depicted by arrows. The crystal structure sketch is the same as in
Fig.~\ref{F-str}.}
\end{figure}

In conclusion, DFT provides a robust qualitative microscopic magnetic
model and unambiguously yields the signs of the leading $J_i$'s, as
revealed by excellent agreement with the experimentally observed
magnetic structure (Fig.~\ref{F-model}).  However, the accuracy of the
individual magnetic couplings is not enough to provide a quantitative
agreement with the experiments. In the next step, the evaluated model is
refined by a simulation of its magnetic properties and subsequent
comparison to the experiments.

\subsection{\label{S-simul}Simulations}
Since the microscopic magnetic model is 2D and frustrated, numerically
efficient techniques that account for the thermodynamic limit, such as
quantum Monte Carlo (QMC) or density-matrix renormalization group
(DMRG), can not be applied. Although new methods are presently under
active development,\cite{[{See, e.g., }] [{}] MERA, *PEPS} a standard
method to address such problems is exact diagonalization (ED) of the
respective Hamiltonian matrix performed on finite lattices of $N$
spins.\cite{2D_ED_overview} However, the performance of present-day
computational facilities limits the size of feasible finite lattices to
$N$\,=\,24 sites\cite{kagome_thermodyn_herb,*volb_DFT} for
finite-temperature properties or $N$\,=\,42
sites\cite{ED_N42_star_lattice, *kagome_GS_ED_N42_gap_AML} for the GS.
Additional restrictions arise from topological features of the
particular model, e.g., only certain values of $N$ fit to periodic
boundary conditions.  Here, we apply ED to study the magnetic properties
of \cucd\ using exact diagonalization on $N$\,=\,16 sites (for the
magnetic susceptibility) and $N$\,=\,32 sites (for GS spin-spin
correlations and the GS magnetization process in magnetic field).

Due to the ambiguous choice of the DCC and $U_{3d}$, our DFT
calculations do not provide a precise position of \cucd\ in the
parameter space of the proposed \jd--\jti--\jtii--\jit\ model. However,
the ratios of the leading couplings (Table~\ref{T-tJ}, three bottom
lines) follow a distinct trend: the AMF solutions yield larger
\jti\,:\,\jd\ and substantially smaller $|\jit|$\,:\,\jd\ values, than
FLL. Compared to the difference between the AMF and FLL solutions, the
value of $U_{3d}$ and the functional used influence the ratios only
slightly. Therefore, we adopt the two different DFT+$U$ solutions: the
AMF solution \jd\,:\,\jti\,:\,\jtii\,:\,\jit\ $\simeq$
1\,:\,0.20\,:\,0.45\,:\,$-$0.30 (AMF, $U_{3d}$\,=\,6.5\,eV) and the FLL
solution \jd\,:\,\jti\,:\,\jtii\,:\,\jit\ $\simeq$
1\,:\,0.15\,:\,0.25\,:\,$-$0.70 (FLL, $U_{3d}$\,=\,9.5\,eV) as starting
points for simulations of magnetic susceptibility.

\begin{figure}[tb]
\includegraphics[width=8.6cm]{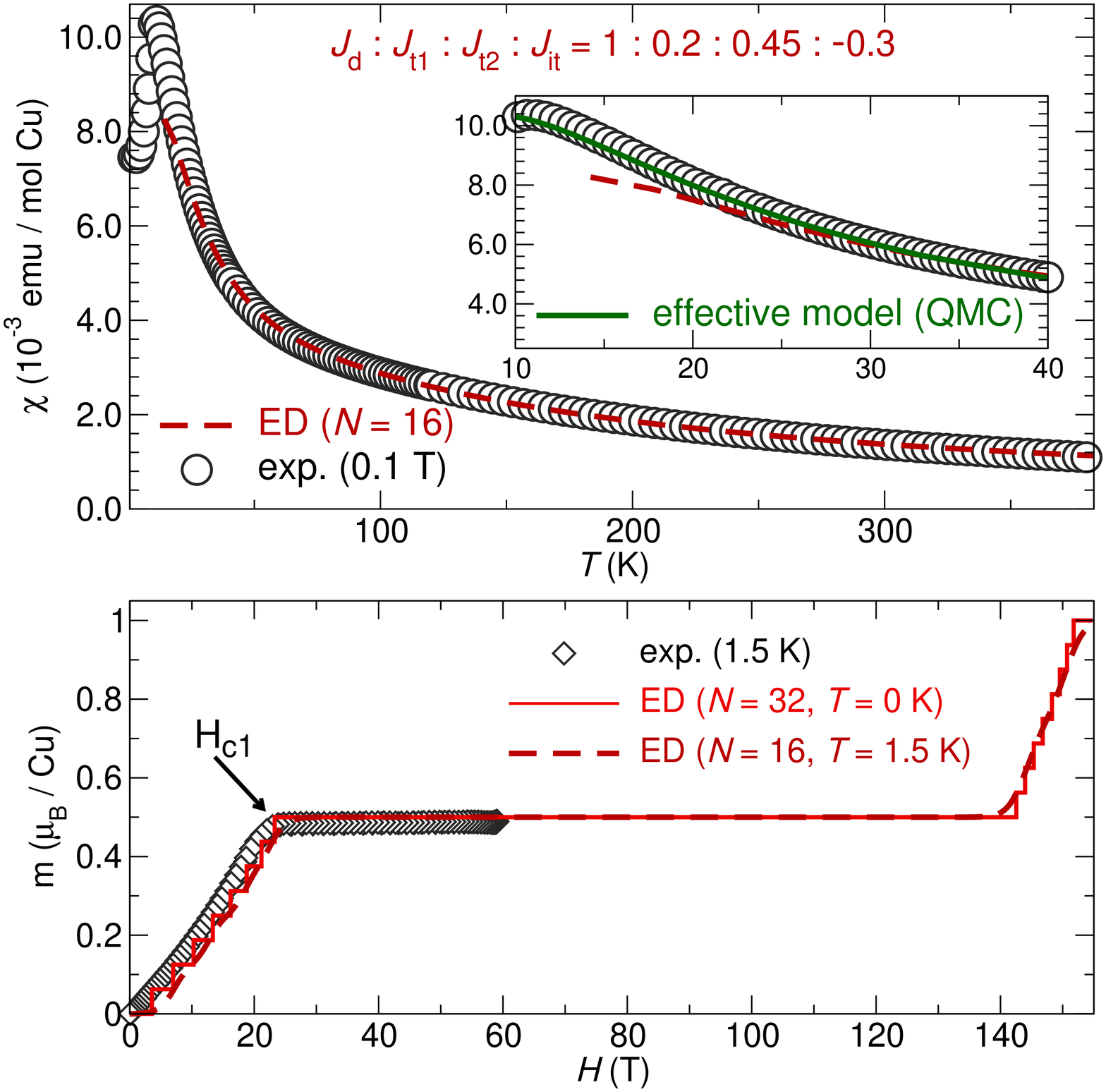}
\caption{\label{F-simul}(Color online) Top panel: ED ($N$\,=\,16 sites)
fit to the experimental (exp.) magnetic susceptibility. Deviations at
$T$\,$<$\,30\,K are an extrinsic finite-size effect. Inset: fit to the
low-temperature $\chi(T)$ using QMC for the effective model
(Sec.~\ref{S-PT}) on a $N$\,=24$\times$24 sites finite lattice. Note the
improved agreement between 15 and 30\,K, compared to the ED fit.  Bottom
panel: magnetization curve measured at 1.5\,K and simulated by ED
(finite-temperature magnetization for $N$\,=\,16 and the GS
magnetization for $N$\,=\,32 sites) scaled using the parameters (\jd\
and $g$-factor) obtained from the fit to $\chi(T)$ (Table~\ref{T-fit},
top line).}
\end{figure}

We have simulated the magnetic susceptibility on a $N$\,=\,16 sites
finite lattice (Fig.~\ref{F-sisj}, bottom) with periodic boundary
conditions, and by keeping the \jd\,:\,\jti\,:\,\jtii\,:\,\jit\ ratios
fixed. Such simulations yield the reduced magnetic susceptibility
$\chi^*(T^*)$, which can be compared to the experimental $\chi(T)$ by
fitting free parameters: the overall energy scale \jd, the $g$-factor
and the temperature-independent contribution $\chi_0$.\footnote{For the
fitting, we used Eq.~2 from Ref.~\onlinecite{HC_dio_DFT_QMC_chiT}.
$C_{\text{imp}}$ was set to zero.} In
accord with the Mermin--Wagner theorem,\cite{Mermin_Wagner} our 2D
magnetic model does not account for long-range magnetic ordering, hence
the simulated magnetic susceptibility is valid only for
$T$\,$>$\,$T_{\text{N}}$\,=\,9.8\,K.\cite{CC_Cu2CdB2O6_NPD_MH} Moreover,
at low temperatures (even above $T_{\text{N}}$) finite-size effects become
relevant.  To find an appropriate temperature range, we used an
auxiliary eight-order high-temperature series
expansion,\cite{HTSE_Richter} and thus evaluated 30\,K as a reasonable
lowest fitting temperature for the $N$=\,16 sites finite lattice.

The fitted values of \jd, $g$, and $\chi_0$ are summarized in
Table~\ref{T-fit}. Since the FLL solution (Table~\ref{T-fit}, bottom
line) yields unrealistically large $g$\,=\,2.457, it can be safely ruled
out. By contrast, the AMF solution (Table~\ref{T-fit}, first row) yields a
reasonable value of $g$\,=\,2.175.\cite{Cu2P2O7_DFT_chiT_MH,
*Cu2V2O7_DFT, *Cu2As2O7} The resulting fit to the experimental
$\chi(T)$ curve is shown in Fig.~\ref{F-simul} (top). 

\begin{table}[b]
\caption{\label{T-fit} 
Fitted values of \jd\ (in K), the $g$-factor and $\chi_0$ (in
$10^{-4}\times$\,emu\,$[$mol Cu$]^{-1}$), as well as the critical field
$H_{\text{c}1}$ (in T), corresponding to the onset
of a magnetization plateau. The simulations have been performed on
$N$\,=\,16 sites finite lattices using ED.}
\begin{ruledtabular}
\begin{tabular}{l c c r r}
\jd\,:\,\jti\,:\,\jtii\,:\,\jit & \jd & $g$ & $\chi_0$ &
$H_{\text{c}1}$ \\ \hline
1\,:\,0.20\,:\,0.45\,:\,$-0.30$ (AMF) & 178.0 & 2.175 & 0.9 & 24.1 \\
1\,:\,0.15\,:\,0.25\,:\,$-0.70$ (FLL) & 291.9 & 2.457 & $-$2.9 & 23.7 \\
\end{tabular}
\end{ruledtabular}
\end{table}

Next, we challenge the solutions from Table~\ref{T-fit} using the
high-field magnetization data (Fig.~\ref{F-simul}, bottom). In
particular, the critical field $H_{\text{c}1}$, corresponding to the
onset of the magnetization plateau (Fig.~\ref{F-simul}, bottom), can be
estimated independently for both experimental and simulated curves.
Experimentally, $H_{\text{c}1}$\,=\,23.2\,T was evaluated by a
pronounced minimum in $\partial^2m/\partial{}H^2$.\footnote{The
simulated magnetization curve exhibits a step-wise behavior which
originates from the finite size of the lattice and the low temperature
$T$/\jd\,=\,8.4$\times$10$^{-3}$ used for the ED simulation. } The
simulated magnetization curves $m(h^{*})$, where $h^{*}$ is the reduced
magnetic field in the units of \jd, were scaled to the experimental data using
the expression
\be
H=\left(k_{\text{B}}\,\jd\,g^{-1}\,\mu_{\text{B}}^{-1}\right)h^{*},
\ee
and by adopting the values of \jd\ and $g$ from the fits to $\chi(T)$.
This way, $H_{\text{c}1}$ can be estimated directly, i.e., without using
additional parameters beyond \jd\ and $g$. Such $H_{\text{c}1}$
estimates are given in the last column of Table~\ref{T-fit}. Both the
AMF and the FLL solutions are in excellent agreement with the
experimental $H_{\text{c}1}$.
Therefore, only the AMF solution \jd\,:\,\jti\,:\,\jtii\,:\,\jit\
$\simeq$ 1\,:\,0.20\,:\,0.45\,:\,$-$0.30 yields good agreement for both
the $g$-factor and the $H_{\text{c}1}$ values. Thus, in the following,
we restrict ourselves to the analysis of this solution
(Table~\ref{T-fit}, first row).

So far, the simulations evidence a good agreement between the DFT-based
model and the experimental information on thermodynamical properties.
In the following, we perform an elaborate theoretical study of the
proposed model. A deep insight into its physics enables further
experimental verification of our microscopic scenario and allows to get
prospects for new experiments.

\begin{figure}[tb]
\includegraphics[width=0.95\linewidth]{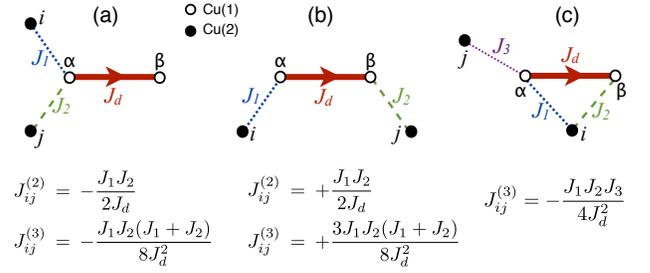}
\caption{\label{fig:2ndAND3rdOrderProcesses} 
(Color online) The three topologically different clusters that contribute 
up to third order in the effective exchange $J_{ij}$ in the limit $J_{1,2,3}\!\ll\!\jd$,  
where $J_{1}$, $J_{2}$, and $J_3$ belong to the set: \jti, \jtii, or \jit.  
Here $J_{ij}^{(2)}$ and $J_{ij}^{(3)}$ denote, respectively, the contributions from 
second and third-order perturbation theory.  For any given cluster, we
include only processes that involve each of the weak couplings at least
once.}
\end{figure}

\begin{figure*}[t]
\includegraphics[width=\linewidth]{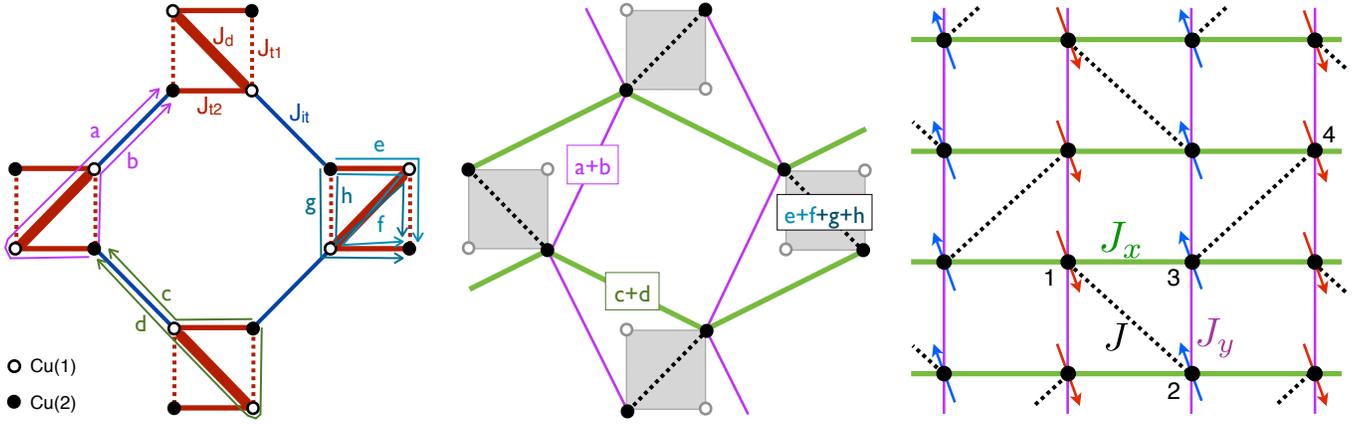}
\caption{\label{fig:PertTheory}(Color online) The effective interactions
between the Cu(2) sites (filled circles) up to lowest second order.  Left panel:
Eight paths $a-h$ that provide a nonzero exchange interaction up to second order. 
Middle panel: effective model based on $a$---$h$ paths:
$J_x=c+d$, $J_y=a+b$, and $J=e+f+g+h$. In third order, one should also include the processes from the third cluster of Fig.~\ref{fig:2ndAND3rdOrderProcesses}, apart from a renormalization of the amplitudes $a$---$h$ (see text). 
Right panel: The resulting effective model in the topologically
equivalent square lattice version of the structure. The indices 1-4
label the four sites of the unit cell of this lattice model.
The arrows denote the classical GS of the model. }
\end{figure*}

\section{\label{S-model-prop}Properties of the magnetic model}

\subsection{\label{S-PT}Effective low-energy theory}\label{Sec:EffTheory}
From the above discussion, it has become clear that the intradimer
exchange \jd\ is considerably larger than the remaining energy scales in
the problem.  This separation of energy scales leads naturally to the
idea that the physics of the problem can be understood by a perturbative
expansion around the limit \jti\,=\,\jtii\,=\,\jit\,=\,0.  In this limit, the
Cu(1) sites coupled by $J_{\text{d}}$ form quantum-mechanical singlets, while the
remaining, Cu(2) sites are free to point up or down, which defines a
highly degenerate GS manifold.  By turning on the remaining couplings
\jti, \jtii, and \jit, the Cu(2) sites begin to interact with each other
through the virtual excitations of the Cu(1) dimers out of their singlet
GS manifold.  By integrating out these fluctuations, one can derive an
effective low-energy model for the Cu(2) sites only.  As we 
show below, a degenerate perturbation theory up to third order gives a 
remarkably accurate description of the low-energy physics of the
problem.

Figure \ref{fig:2ndAND3rdOrderProcesses} shows the three topologically
different clusters $G$ that involve a given Cu(1) dimer and contribute up to
third order in perturbation theory. To avoid double counting of the
processes that live on subclusters of $G$,\cite{Oitmaa} we keep only those
processes that invoke each weak coupling of $G$ at least once.   Here the weak
couplings $J_{1,2,3}\!\ll\!\jd$ belong to the set: \jti, \jtii, and \jit.   

For the following discussion, it is essential to emphasize the difference
between the first two clusters in Fig.~\ref{fig:2ndAND3rdOrderProcesses}:  in
(a) the two couplings involve the same site $\alpha$ of the dimer, while in (b)
they involve both sites $\alpha$ and $\beta$. Since the singlet wave function
$|s\rangle_{\alpha\beta}=\frac{1}{\sqrt{2}}\left(
|\!\uparrow_\alpha\downarrow_\beta\rangle
-|\!\downarrow_\alpha\uparrow_\beta\rangle \right)$ on the strong dimer is
antisymmetric with respect to interchanging $\alpha$ and $\beta$, the two
processes have opposite sign (and are equal in magnitude up to second order).
In particular, if $J_1 J_2>0$, the effective coupling between the sites $i$ and
$j$ is ferromagnetic in (a) but antiferromagnetic in (b).  For the third
cluster, we have a third-order process where each of the three weak couplings
$J_1$, $J_2$ and $J_3$ has been involved once in the whole process.

Using these quite general results, we may turn to our spin lattice
and examine the possible effective processes that are active up to third order.  
We first discuss the contributions of the type shown in Fig.~\ref{fig:2ndAND3rdOrderProcesses}(a) and (b).    
Altogether, we find eight different paths $a$--$h$ $[$these are shown in Fig.
\ref{fig:PertTheory}(a)$]$ with the following amplitudes:
\bea
&& a=+\frac{\jtii \jit}{2 J_{\text{d}}} \left(1+\frac{3(\jtii+\jit)}{4J_{\text{d}}} \right) = -\frac{2403}{32000}J_{\text{d}}, \nonumber\\
&& b=-\frac{\jti \jit}{2 J_{\text{d}}} \left(1+\frac{\jti+\jit}{4J_{\text{d}}} \right) = +\frac{117}{4000}J_{\text{d}}, \nonumber\\
&& c=-\frac{\jtii \jit}{2 J_{\text{d}}} \left(1+\frac{\jtii+\jit}{4J_{\text{d}}} \right) = +\frac{2241}{32000}J_{\text{d}} , \nonumber\\
&& d=+\frac{\jti \jit}{2 J_{\text{d}}} \left(1+\frac{3(\jti+\jit)}{4J_{\text{d}}} \right) = -\frac{111}{4000}J_{\text{d}}, \nonumber\\
&& e=g=-\frac{\jti \jtii}{2 J_{\text{d}}} \left(1+\frac{\jti+\jtii}{4J_{\text{d}}}\right) = -\frac{837}{16000}J_{\text{d}}, \nonumber\\
&& f= +\frac{\jtii^2}{2 J_{\text{d}}} \left(1+\frac{3\jtii}{2J_{\text{d}}}\right) = \frac{5427}{32000}J_{\text{d}} ,\nonumber\\
&& h= +\frac{\jti^2}{2 J_{\text{d}}} \left(1+\frac{3\jti}{2J_{\text{d}}}\right)= \frac{13}{500}J_{\text{d}}~.\nonumber
\eea
In the right-hand side equalities, we made use of the estimated ratios
of the original couplings from Table~\ref{T-fit} (the AMF solution).
Including now the contributions of the type shown in
Fig.~\ref{fig:2ndAND3rdOrderProcesses}(c), we get altogether three
different effective exchange interactions between the Cu(2) spins: 
\bea
&& J_x = c+d -\frac{\jti\jtii\jit}{2J_{\text{d}}^2} \simeq +0.05578 J_{\text{d}}  ~,\label{eqn:Jx}\\ 
&& J_y =a+b -\frac{\jti\jtii\jit}{2J_{\text{d}}^2}  \simeq -0.03234 J_{\text{d}} ~,\label{eqn:Jy}\\ 
&& J = e+f+g+h-\frac{\jti\jtii(\jti+\jtii)}{2J_{\text{d}}^2} \simeq 0.06172 J_{\text{d}}~.\label{eqn:J} 
\eea
These couplings are shown in the right panel of
Fig.~\ref{fig:PertTheory}, where we use the square-lattice
representation that is topologically equivalent to our effective model.
Hence, the resulting effective model is the anisotropic
Shastry-Sutherland lattice model with an AFM exchange coupling $J$ for
the diagonal bonds, an AFM exchange $J_x$ for the horizontal bonds, and
an FM exchange $J_y$ for the vertical bonds. This model readily explains
the observed magnetic ordering on the Cu(2) sites since,  on
the classical level, one can minimize all interactions simultaneously in
the collinear ``stripe'' configuration shown in the right panel of Fig.
\ref{fig:PertTheory}.

To challenge the applicability of the effective model, we refer again to
the experimental $\chi(T)$ dependence. The nonfrustrated nature of the
effective model enables us to use numerically efficient QMC techniques.\cite{ALPS}
To this end, we have simulated $\chi^*(T^*)$ of the effective
model on a finite $N$\,=\,24$\times$24 sites lattice with periodic boundary
conditions.  A parameter-free fit to the experimental curve is
obtained by using Eqs. (\ref{eqn:Jx})-(\ref{eqn:J}) above, and
the values of \jd\ and $g$ from the ED fit (Table~\ref{T-fit}). By
construction, the effective model is valid at low temperatures
($T$\,$\ll$\,\jd), but above the long-range magnetic ordering temperature
$T_N$. In this temperature range, the effective model yields excellent
agreement with the experimental $\chi(T)$ (see inset in top panel of
Fig.~\ref{F-simul}), justifying it as an appropriate low-energy model for
\cucd.

\subsection{The size of the Cu(2) moments: linear spin wave theory in the
effective model and QMC simulations} 
The size of the Cu(2) moment in the GS of the system is not equal to the
classical value $g\mu_B/2$, but it will be somewhat lower due to quantum
fluctuations. A similar correction is expected, e.g., for the GS energy.
We can calculate these corrections by a standard $1/S$ semiclassical
expansion around the stripe phase keeping only the quadratic portion of
the fluctuations.  The details of this calculation are provided in
Appendix \ref{App:LSWT}, and here we only discuss the main results.

\begin{figure}[!t]
\includegraphics[width=0.9\linewidth]{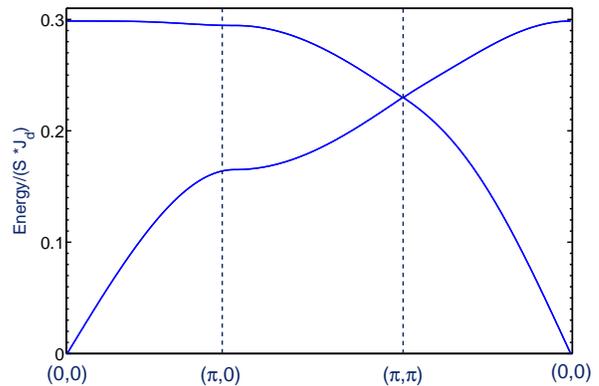}
\caption{\label{fig:LSWTdispersions}(Color online) Magnon dispersions (along
certain symmetry directions in the BZ) obtained from linear spin wave theory
around the stripe phase of the Cu(2) spins (see right panel of Fig.
\ref{fig:PertTheory}).  There are four magnon branches but these pair up into
two 2-fold degenerate branches, which is due to the symmetry of interchanging
the sites (1,2) with (3,4) of the unit cell (see right panel of Fig.
\ref{fig:PertTheory}).}
\end{figure}

Figure \ref{fig:LSWTdispersions} shows the magnon dispersions of the effective
model along certain symmetry directions in the Brillouin zone (BZ).  There are
four magnon branches but these pair up into two 2-fold degenerate branches due
to the symmetry of interchanging the sites (1,2) with (3,4) in the unit cell
(see right panel of Fig. \ref{fig:PertTheory}). The accuracy of the magnon
dispersions was cross-checked by the \textsc{McPhase} code.\cite{McPhase}

Let us now look at the quadratic correction to the GS energy which, as shown in
Appendix \ref{App:LSWT}, consists of two terms, $\delta E_1$ and $\delta E_2$.  The
first term is given by $\delta E_1 = -2 \xi N_{uc} S$, where $\xi \equiv J + 2 J_x -
2 J_y$, and $N_{uc}$ is the number of unit cells.  So adding $\delta E_1$ to
the classical GS energy $E_{\text{class}} = -2 \xi N_{uc} S^2$, amounts to a
renormalization of the spin from $S$ to $\sqrt{S(S+1)}$.  The second
quadratic correction $\delta E_2$ stands for the zero-point energy of the
final, decoupled harmonic oscillators of the theory $[$see
Eq.~(\ref{eqn:dE2})$]$, and can be calculated by a numerical integration
over the BZ. Altogether, we find
\bea 
E_{\text{class}}/N_{uc} &=& -0.119~J_{\text{d}} \\
\delta E_1 /N_{uc} &=& -0.238 ~J_{\text{d}}\\
\delta E_2 /N_{uc}  &\simeq& + 0.2092 ~J_{\text{d}}. 
\eea 
Thus, quadratic fluctuations actually reduce the GS energy by about $24\%$.  We
now turn to the renormalization of the magnetic moment, $\delta S$.  As
explained in Appendix~\ref{App:LSWT}, this can also be calculated by a
numerical integration over the BZ, see Eq. (\ref{eqn:dSz}).  We find  
\be
\delta S \simeq 0.159 \Rightarrow S = 0.5 -\delta S \simeq 0.341 ~.
\ee
With $g\simeq 2.175$ we get a local magnetic moment on the Cu(2) sites of 
$m \simeq 0.742 ~\mu_B$, in line with the experimental value of $m \simeq 0.83
~\mu_B$.\cite{CC_Cu2CdB2O6_NPD_MH}

\begin{figure}[tb]
\includegraphics[width=8.6cm]{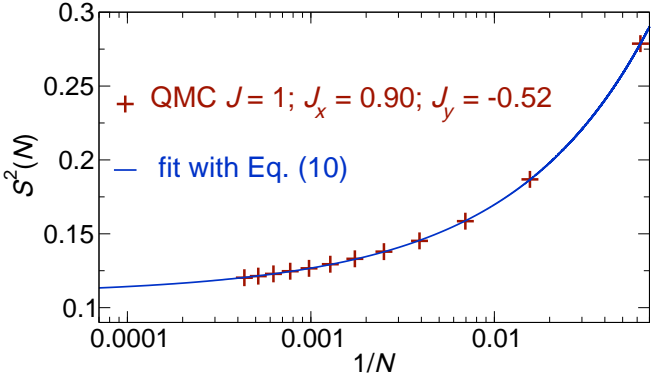}
\caption{\label{fig:QMCeff}(Color online) 
Finite-size extrapolation of the quantity $S^2$($N$) defined in
Eq.~(\ref{eqn:mSqN}). Finite-$N$ values (crosses) were evaluated using
QMC simulations for the effective model of the Cu(2) spins for a series
of finite lattices up to $N$\,=\,48$\times$48 sites.  The fit
to Eq. (\ref{eqn:mSqN-scaling}) yields $\sigma_1$\,=\,0.5509,
$\sigma_2$\,=\,0.6436, $\sigma_3$\,=\,$-0.5097$, and
$S_{\infty}$\,$\simeq$\,0.3297, with the extrapolated magnetic moment
$m$\,=\,$g\mu_{B}S_{\infty}$\,$\simeq$\,0.717\,$\mu_{\text{B}}$.}
\end{figure}

Since the model is not frustrated, we expect that the above prediction
from linear spin wave theory is a good approximation.  To verify this
conjecture, we simulate the static structure factor $\mathbb{S}$ corresponding
to the propagation vector of the collinear state ($\pi$,0) using QMC, which
readily yields the order parameter $S$ $[$Eq.~(33) in
Ref.~\onlinecite{SL_QMC_finite_size_scaling}$]$:
\be\label{eqn:mSqN}
S^2(N)=\frac{3\mathbb{S}_{(\pi,0)}}{N}.
\ee
The results are presented in Fig.~\ref{fig:QMCeff}.  We evaluated
finite 2D lattices with up to $N$\,=\,48$\times$48 sites, and used
Eq.~(39b) from Ref.~\onlinecite{SL_QMC_finite_size_scaling} for
the finite-size scaling (Fig.~\ref{fig:QMCeff}):
\be\label{eqn:mSqN-scaling}
S^2(N)=S^2_{\infty}+\frac{\sigma_1}{\sqrt{N}} +
\frac{\sigma_2}{N} +
\frac{\sigma_3}{\sqrt{N}^3},
\ee
where $S_{\infty}$ is the magnetic moment, extrapolated to the
thermodynamic limit.  The resulting $S_{\infty}$\,$\simeq$\,0.3297
yields $m$\,=\,$g\mu_{B}S_{\infty}$\,$\simeq$\,0.717\,$\mu_B$, corroborating the
linear spin-wave theory result.

\subsection{Staggered polarization on the Cu(1) dimers}\label{Sec:StaggeredMoment}
\begin{figure}[!t]
\includegraphics[width=0.8\linewidth]{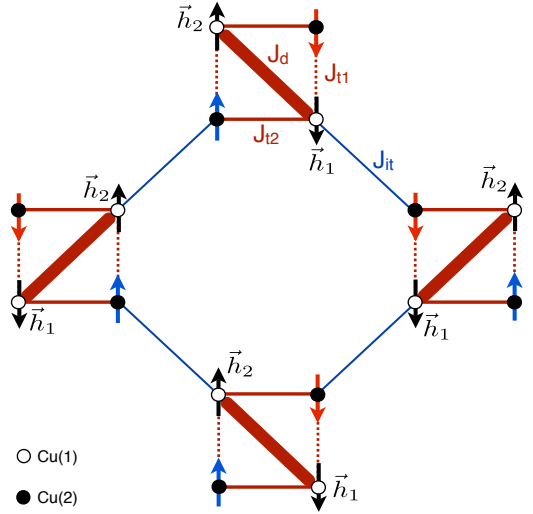}
\caption{\label{fig:StripePhase}(Color online) The classical ground
state of the effective model derived in Sec. \ref{Sec:EffTheory}.  The
vectors $\vec{h}_{1,2}$ denote the local exchange 
fields on the Cu(1) sites exerted by the magnetically ordered Cu(2)
sites (blue and red arrows).}
\end{figure}

The effective low-energy theory derived above in Sec. \ref{Sec:EffTheory} is a
theory for the projection of the full wave function of the system onto the
manifold where all Cu(1) dimers form singlets. However, there is also a
finite GS component out of this manifold. In particular, once the Cu(2) sites order
magnetically in the collinear stripe phase (see right panel of Fig.
\ref{fig:PertTheory}), they will exert a finite exchange field on the Cu(1)
dimers. A simple inspection of Fig. \ref{fig:StripePhase} tells us that
this field is staggered, i.e., the local fields on the two Cu(1) spins 
of any given dimer are opposite to each other.  Specifically, the
fields $\vec{h}_{1,2}$, as shown in Fig. \ref{fig:StripePhase}, are given by
\be\label{eqn:LocalField}
\vec{h}_{1,2}\!=\!\pm\left(\jti - \jtii + \jit\right)\!
\langle\vec{S}_{\text{Cu}_2}\rangle = \mp 0.55 J_{\text{d}} \langle\vec{S}_{\text{Cu}_2}\rangle, 
\ee
where the last equation follows again from the DFT-based estimates for
the ratios of the leading couplings (Table~\ref{T-fit}, first row). 

Now, the important point is that the staggered field does not commute
with the exchange interaction on the dimer, and as a result it can polarize the
system immediately. This is very different from the case of a uniform field,
where one must exceed the singlet-triplet gap $J_{\text{d}}$ in order to
(uniformly) polarize the sites of the dimer.

To assess the amount of the induced polarization on the Cu(1) dimers, we
consider a single AFM dimer with the exchange \jd\ in the presence of a
staggered field $h$, described by the Hamiltonian \be
\mc{H} = J_{\text{d}} ~\vec{S}_1 \cdot \vec{S}_2 -h (S_1^z - S_2^z)
\ee  
The eigenstates of the dimer for $h=0$ are the singlet GS,  
$|s\rangle= \frac{1}{\sqrt{2}} \left( |\!\uparrow\downarrow\rangle-|\!\downarrow\uparrow\rangle \right)$, 
and the triplet excited states 
$|t_1\rangle=|\!\uparrow\uparrow\rangle$,  $|t_{-1}\rangle = |\!\downarrow\downarrow\rangle$, and 
$|t_0\rangle = \frac{1}{\sqrt{2}} \left( |\!\uparrow\downarrow\rangle + |\!\downarrow\uparrow\rangle \right)$. 
The staggered field leaves $|t_1\rangle$ and $|t_{-1}\rangle$ intact, namely $\mc{H} |t_{\pm 1}\rangle = \frac{J}{4} |t_{\pm 1}\rangle$,  
but it mixes $|s\rangle$ with $|t_0\rangle$ as follows       
\bea
&&\mc{H} |t_0\rangle = \frac{J_{\text{d}}}{4} |t_0\rangle -h |s\rangle\\
&&\mc{H} |s\rangle = -\frac{3J_{\text{d}}}{4} |s\rangle -h |t_0\rangle ~.
\eea
A straightforward diagonalization of the Hamiltonian in the subspace
of $\{|s\rangle, |t_0\rangle\}$ leads to the eigenvalues
\be
E_\pm = -\frac{J_{\text{d}}}{4} \pm \sqrt{\left( \frac{J_{\text{d}}}{2} \right)^2+h^2} ~.
\ee
For $h$\,=\,0, these results reduce to the energies $J_{\text{d}}/4$ and
$-3J_{\text{d}}/4$ as expected. The GS is given by 
\be
|\psi_-\rangle  = \frac{1}{N} \left[ \left(J_{\text{d}}/2+\sqrt{J_{\text{d}}^2/4+h^2}\right)  |s\rangle + h|t_0\rangle  \right] ~,
\ee
where $N^2 = 2h^2 + J_{\text{d}}^2/2 + J_{\text{d}} \sqrt{J_{\text{d}}^2/4+h^2}$. As expected,  $|\psi_-\rangle\!\mapsto\!|s\rangle$  for $h\!\mapsto\!0$, while for 
$h \gg J_{\text{d}}$, $|\psi_-\rangle\!\mapsto |\!\uparrow\downarrow\rangle$.

We may also look at the GS expectation value of the local polarizations on the dimer. We find 
\be\label{eqn:S12z}
\langle \psi_-| S_{1,2}^z | \psi_- \rangle  = \pm  \frac{h
(J_{\text{d}}/2+\sqrt{J_{\text{d}}^2/4+h^2})}{2h^2+J_{\text{d}}^2/2+J_{\text{d}}
\sqrt{J_{\text{d}}^2/4+h^2}}~.
\ee
For $h=0$ we recover $\langle \psi_-| S_{1,2}^z | \psi_- \rangle=0$,
while for $h\gg J_{\text{d}}$ we get maximum polarizations $\langle \psi_-|
S_{1,2}^z | \psi_- \rangle \mapsto \pm 1/2$, which corresponds to the state $|\!\!\uparrow\downarrow\rangle$. 
In the linear response regime $h\!\!\ll\!\!J_{\text{d}}$, we get 
$\langle\psi_-|S_{1,2}^z|\psi_-\rangle \simeq \pm h/J_{\text{d}}$, 
thus the staggered susceptibility is $\chi_s = 1/J_{\text{d}}$.

We can now apply these results to our case. Using Eqs.~(\ref{eqn:LocalField}) and (\ref{eqn:S12z}), 
the polarization on the Cu(1) sites can be estimated as 
\be
|\langle S_{\text{Cu}_1}\rangle| \simeq 0.176 \simeq 0.516 |\langle S_{\text{Cu}_2}\rangle|~, 
\ee
which is again in very good agreement with the experimental $|\langle
S_{\text{Cu}_1}\rangle|$\,=\,0.54$|\langle S_{\text{Cu}_2}\rangle|$ from
Ref.~\onlinecite{CC_Cu2CdB2O6_NPD_MH}.

\subsection{The 1/2-plateau phase and the critical field $H_{c1}$}\label{Sec:OneHaldPlateau}
The separation of energy scales in \cucd\ is a natural reason for the
formation of the 1/2 magnetization plateau.  Due to the strong
$J_{\text{d}}$, the Cu(1) spins involved in the $J_{\text{d}}$-dimers
are much harder to polarize with a uniform field than the weakly coupled
Cu(2) spins.  Naturally then, the 1/2-plateau corresponds to the state
with the Cu(2) spins fully polarized and the Cu(1) dimers still carrying a
strong singlet amplitude and a finite staggered polarization. 

Here, we go one step further and calculate $H_{c1}$, the onset field of
the 1/2-plateau by considering the one-magnon spectrum of the effective
model in the fully polarized phase.  The one-magnon space consists of
states with all but one spins pointing up.  We label the possible
positions of the down spin by one index for the unit cell and another
index $\alpha=1-4$ which specifies one of the 4 inequivalent Cu sites
within the unit cell (see sites 1-4 in the right panel of Fig.
\ref{fig:PertTheory}).  A straightforward Fourier transform gives the
following $4\times 4$ Hamiltonian matrix in the one-magnon space
\begin{widetext}
\be \nonumber
\delta\mc{H}_{\text{1-magnon}} (\vec{k}) = \frac{1}{2}\left(
\begin{array}{c c c c}
2\lambda & J & J_x (1+e^{-i k_x}) & J_y (e^{-i k_x}+e^{-i (k_x+k_y)}) \\
J & 2\lambda & J_y (1+e^{-i k_y}) & J_x (e^{-i k_y}+e^{-i (k_x+k_y)}) \\
J_x (1+e^{i k_x}) & J_y (1+e^{i k_y})  & 2\lambda & J\\
J_y (e^{i k_x}+e^{i (k_x+k_y)}) & J_x (e^{ i k_y}+e^{ i (k_x+k_y)}) & J & 2\lambda 
\end{array}
\right)~, 
\ee
\end{widetext}
where we measure all energies relative to the energy of the fully
polarized state, and $\lambda=g \mu_B H- \left( J_x+J_y+J/2 \right)$.
Using Eqs.~(\ref{eqn:Jx})-(\ref{eqn:J}), we find that one of the four branches
of the one-magnon spectrum becomes soft (reaches zero energy) at $\vec{k}=0$ when 
$g \mu_B H = J + 2 J_x = \frac{1109}{6400}J_{\text{d}}$. 
This soft mode signals the instability of the fully polarized state, hence
\be
H_{c1} = \frac{1109}{6400} \frac{J_{\text{d}}}{g\mu_B}~.
\ee
With $g$\,$\simeq$\,2.175 and $J_{\text{d}}$\,=\,178\,K (AMF solution
from Table~\ref{T-fit}), we obtain $H_{c1}$\,$\simeq$\,21.11\,T, which
is in very good agreement with the experimental value of
$H_{c1}$\,=\,23.2\,T.  This confirms that the third-order perturbation
theory gives an adequate quantitative description of the low-energy
physics.  It also confirms that our set of exchange parameters gives an
accurate quantitative description of the magnetism in this compound.

\subsection{Exact Diagonalizations}
\begin{figure}[tb]
\includegraphics[width=8.6cm]{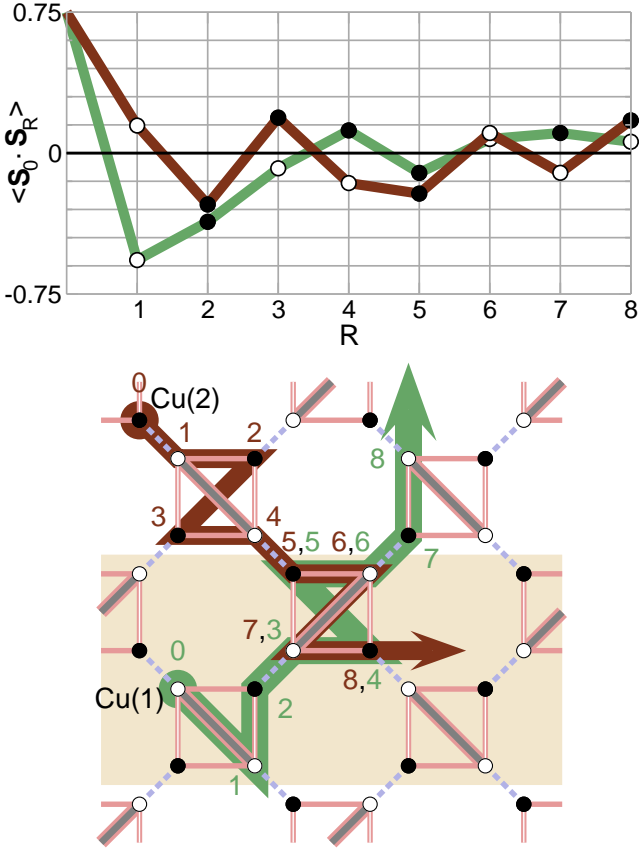}
\caption{\label{F-sisj} (Color online) Upper panel: spin
correlations \mbox{$\langle{}\vec{S}_0\cdot\vec{S}_{\text{R}}\rangle$} in the
magnetic ground state of \cucd, simulated on the 32-site finite lattice
with periodic boundary conditions using ED. Lower panel: $N$\,=\,32
sites (full plot) and $N$\,=\,16 sites (shaded rectangle) finite
lattices used in our ED.  $\vec{S}_0$--$\vec{S}_{\text{R}}$ pathways are
used for the
$\langle{}\mathbf{S}_0\!\cdot\!\mathbf{S}_{\text{R}}\rangle$(R) plot in
the upper panel. For the notation of exchange couplings, see
Fig.~\ref{F-model}. }
 \end{figure}

We are now going to check the above physical picture obtained from the effective
model by numerical exact diagonalizations on the original model.  We first address the
nature of the magnetic GS by evaluating the spin-spin correlations
$\langle{}\bf{S}_0\!\cdot\!\bf{S}_{\text{R}}\rangle$ {as a function of the
``distance'' R} (Fig.~\ref{F-sisj}).  In order to elucidate the
difference in the coupling regimes of Cu(1) and Cu(2) spins, we choose
two independent paths on the 2D spin lattice, with different initial
spins $\bf{S}_0$ (Fig.~\ref{F-sisj}, lower panel). As expected for
strong \jd, the structural dimers exhibit strong AFM correlation
$\langle{}\bf{S}_0\!\cdot\!\bf{S}_{\text{1}}\rangle$\,=\,$-0.57$ (green line
in Fig.~\ref{F-sisj}), while the \jit\ bonds bear much weaker FM
correlation $\langle{}\bf{S}_0 \cdot \bf{S}_{\text{1}}\rangle$\,=\,$0.15$
(brown line in Fig.~\ref{F-sisj}).

A rough estimate for the ordered magnetic moment can be obtained from
the spin correlations of the maximally separated spins on a finite
lattice.  The small size of the 32-site finite lattice impedes a direct
evaluation of the order parameters.
However, the same-sublattice correlations for the Cu(1) and Cu(2) 
sublattices, evaluated using the same $N$\,=\,32 cluster,
can be compared to each other. In this way, finite size effects, which
are crucial for the values of $m_{\text{Cu(1)}}$ and $m_{\text{Cu(2)}}$,
should be largely remedied for the $m_{\text{Cu(1)}}$:$m_{\text{Cu(2)}}$
ratio.

The maximal separation in terms of exchange bonds between the
same-sublattice spins on the 32-site finite lattice is five and
corresponds to $\langle{}\bf{S}_0\!\cdot\!\bf{S}_8\rangle$ correlations
in Fig.~\ref{F-sisj}. ED yields
$\langle{}\bf{S}_0\!\cdot\!\bf{S}_8\rangle[\text{Cu(1)}]$\,=\,0.060 and
$\langle{}\bf{S}_0\!\cdot\!\bf{S}_8\rangle[\text{Cu(2)}]$\,=\,0.174 for
Cu(1) and Cu(2), respectively.  In the simplest picture, the quantity
$\sqrt{\langle{}\bf{S}_0\!\cdot\!\bf{S}_8\rangle}[\text{Cu(1)}]$\,:\,$\sqrt{\langle{}\bf{S}_0\!\cdot\!\bf{S}_8\rangle}[\text{Cu(2)}]$\,=\,0.59
should be close to the ratio of the magnetic moments.  Although the
results may be still affected by finite-size effects, a comparison to
the experimental\cite{CC_Cu2CdB2O6_NPD_MH}
$m_{\text{Cu(1)}}$\,:\,$m_{\text{Cu(2)}}$\,=\,0.45\,:\,0.83\,=\,0.54
reveals surprisingly good agreement between theory and experiment. 

\begin{figure}[tb]
\includegraphics[width=8.6cm]{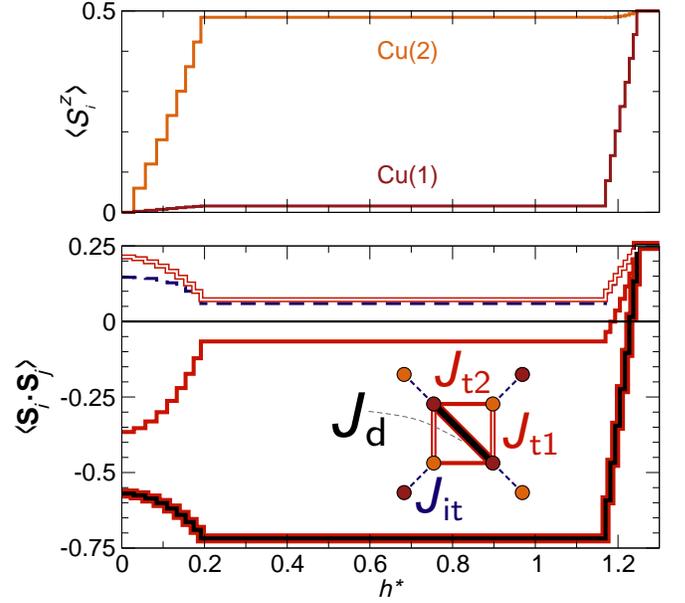}
\caption{\label{F-plateau}(Color online) Magnetization
$\langle{}S^z_i\rangle$ for Cu(1) and Cu(2) spins, as well as spin
correlations $\langle{}\mathbf{S}_i\cdot\mathbf{S}_j\rangle$
corresponding to the leading couplings \jd, \jti, \jtii, and \jit\
(line styles according to the inset) as a function of magnetic field
$h^{*}$ in units of \jd.} \end{figure}

Next, we examine the evolution of the GS expectation values of the local
spin-spin correlations in magnetic field, which are presented in
Fig.~\ref{F-plateau}.  For $h^{*}$\,=\,0, the correlations within the
structural dimers and along the \jtii\ bonds are dominant, amounting to
$-$0.570 and $-$0.366, respectively.  The first one is somewhat weaker than the
pure singlet value of $-$0.75, because of the finite admixture with the triplet
$|t_0\rangle$ in GS, as explained in Sec.~\ref{Sec:StaggeredMoment}. 

Now, according to Fig.~\ref{F-plateau}, a magnetic field enhances the
intradimer correlation.  This counter-intuitive result can be physically
understood as follows. In a finite field, the Cu(2) spins order in the
plane perpendicular to the field (this happens in an infinitesimal
field, since the crystalline anisotropy effects are neglected) and at
the same time develop a uniform component along the field.  While the
former, as discussed in Sec.~\ref{Sec:StaggeredMoment}, gives rise to a
staggered exchange field on the Cu(1) dimers,  the latter induces a
uniform exchange field. By increasing the external field, the staggered
component decreases, while the uniform component increases.  At the
1/2-plateau, the Cu(2) spins are almost fully polarized, hence only the
uniform component survives. Now, in contrast to the staggered component,
the uniform field commutes with the exchange interaction on the Cu(1)
dimer and thus leaves the singlet GS wave function intact. In other
words, the magnetic field suppresses the admixture of a triplet in the
singlet GS.  The steep decrease in the correlation along the \jtii\
bonds in a field reflects the same physics.

At the critical field $h^*$\,$\simeq$\,0.2\ (24\,T), \cucd\ enters the
1/2-plateau phase.  Here, in contrast with the strong AFM correlations within the Cu(1) dimers 
($\langle{}\bf{S}_0\!\cdot\!\bf{S}_{\text{1}}\rangle$\,=\,$-$0.718), the
correlations along the \jti, \jtii\ and \jit\ bonds are much smaller and
similar in size, but have different signs. The negative correlation for
\jtii\ and the positive for \jit\ are in accord with the nature of these
two couplings, AFM and FM, respectively. For weaker AFM \jti, the
correlation is positive due to the frustrated nature of the spin model. 

Since the local spin correlation on the Cu(1) dimers is very close to
the singlet value, the contribution to the total magnetization from the
Cu(1) spins is expected to be negligible.  Therefore, the magnetization
results predominantly from the Cu(2) spins. This picture is supported by
simulations of the local magnetizations $\langle{}S^z_i\rangle$, shown
in the upper panel of Fig.~\ref{F-plateau}.  In the 1/2-plateau phase,
the local magnetization of Cu(2) is
$\langle{}S^z_{\text{Cu2}}\rangle$\,=\,0.483, while the magnetization of
Cu(1) is negligibly small with
$\langle{}S^z_{\text{Cu1}}\rangle$\,=\,0.017, which also agrees with the
main picture obtained from the effective model. Therefore, within the
error from finite-size effects, only about 3\,\% of the Cu(1) moments
are polarized in the 1/2-plateau phase. 

Let us now discuss the width of the 1/2-plateau. This corresponds to the
energy we need to pay to fully polarize a Cu(1) dimer, i.e.\ to convert
a singlet into a triplet excitation. The width of the plateau can be
estimated by neglecting the perturbative corrections that give rise to a
hopping of these triplet excitations (and, thus, to a small dispersion). Our
ED results (upper panel of Fig.~\ref{F-plateau}) agree very well with
this estimate: the critical field at which the 1/2-plateau ends is
$h^*$\,=\,1.17 (143\,T).  

Now, once we pay the energy \jd\ to excite a triplet, we can also
polarize the remaining Cu(1) dimers very quickly due to the small
(perturbative) kinetic energy scale of the triplet excitations.  Indeed,
our ED results for the magnetization above $H_{c2}$ show a rapid
increase up to the saturation value.  In addition, we find no indication
for any other magnetization plateaux above 1/2, which suggests that the
interaction between the triplet excitations is negligible.

\section{\label{S-sum}Summary and outlook} 
The magnetism of \cucd\ was investigated on a microscopic level by using
extensive DFT band structure calculations, ED and QMC simulations, third-order
perturbation theory, as well as high-field magnetization measurements. In
contrast to the previously suggested ``chains + dimers'' model, we
evaluate a quasi-2D magnetic model with mostly antiferromagnetic
couplings: an intradimer exchange \jd\ and two intra-tetramer
interactions \jti\ and \jtii, as well as a ferromagnetic coupling \jit\
between the tetramers. The \jd\,:\,\jti\,:\,\jtii\,:\,\jit\ ratio is
close to 1\,:\,0.20\,:\,0.45\,:\,$-0.30$ with a dominant \jd\ of about
178\,K.  Topologically, this microscopic model can be regarded as
decorated anisotropic Shastry-Sutherland lattice (Fig.~\ref{F-ShSu}).

While the spin pairs forming the structural dimers show a clear tendency
to form magnetic singlets on the \jd\ bonds, the low-temperature
magnetic properties of \cucd\ are governed by interdimer couplings. In
contrast to the strongly frustrated \scbo\ featuring a singlet ground
state, the inequivalence of \jti\ and \jtii\ largely lifts the
frustration and triggers long-range antiferromagnetic ordering. This 
model correctly reproduces the substantial difference between Cu(1) and
Cu(2) ordered moments, as observed experimentally. Our simulations and
the subsequent fitting of the magnetic susceptibility and magnetization
further support the microscopic model and the proposed parameterization. 

It is interesting to compare the key results for \cucd\ with the
archetype Shastry-Sutherland $S$\,=\,$\frac12$ system
SrCu$_2$(BO$_3$)$_2$.  Although the substitution of Sr by Cd
significantly alters the crystal structure and stabilizes a different
magnetic ground state, the microscopic magnetic models of both
systems bear apparent similarities (Fig.~\ref{F-ShSu}). Moreover, the
dominance of the \jd\ coupling in \cucd\ leads to an effective
nonfrustrated model for the Cu(2) spins, which is topologically
equivalent to an anisotropic Shastry-Sutherland lattice model. It is
noteworthy that the phase diagram of such a model was recently
discussed.\cite{CuClLaNb2O7_eff_model_simul} 

One of the main objectives of this study is to stimulate further
experimental activity on \cucd. Such studies are necessary to refine the
parameters of the suggested microscopic model and address yet unresolved
issues.  Our extensive analysis of the magnetic properties of this
complex system delivers several results that should be challenged
experimentally. For example, we have resolved that the Cu(1) dimers
develop a sizable staggered polarization which is suppressed by an
applied field and vanishes at the 1/2 plateau.  This behavior could be
addressed experimentally by nuclear magnetic resonance.\cite{[{See,
e.g.\ }] [{}] SCBO_NMR_stagger, *SCBO_NMR_stagger_DM} On the other hand,
inelastic neutron scattering experiments could challenge the magnon
dispersions calculated for the effective low-energy model. In addition,
the magnetic couplings, associated with edge-sharing and corner-sharing
connections of CuO$_4$ plaquettes, are strongly dependent on Cu--O--Cu
angles.  Therefore, the magnetic properties of \cucd\ could be tuned by
applying external pressure.  Similar experiments on \scbo\
(Ref.~\onlinecite{SCBO_NMR_pressure}) reveal excellent potential of such
studies.

\section*{Acknowledgments}
We thank M. Hase for providing us with the structural data and sharing
his preprint prior to publication. We are also very grateful to M. Rotter, M.
Horvati{\'{c}}, M. Baenitz, R.  Sarkar, and R. Nath for fruitful discussions
and valuable comments.  The high-field magnetization measurements were
supported by EuroMagNET II under the EC contract 228043. A. A. T. acknowledges
the funding from Alexander von Humboldt Foundation.

\appendix
\section{Linear spin wave theory around the stripe phase of the
effective model}\label{App:LSWT} Here, we provide the details of the
linear spin wave theory around the stripe phase of the Cu(2) spins.  We
begin with the effective model Hamiltonian that was derived in Sec.
\ref{Sec:EffTheory} for the Cu(2) spins:
\bea
\mc{H} &\!=\!& \sum_{u} J~\Big( \vec{S}_{1,u} \cdot \vec{S}_{2,u}+\vec{S}_{3,u}\cdot\vec{S}_{4,u} \Big) \nonumber\\
&+& J_x~\vec{S}_{3,u} \cdot \Big( \vec{S}_{1,u} +\vec{S}_{1,u+t_x} \Big) \nonumber\\
&+& J_y~\vec{S}_{3,u} \cdot \Big( \vec{S}_{2,u} +\vec{S}_{2,u+t_y} \Big) \nonumber\\
&+& J_x~\vec{S}_{4,u} \cdot \vec{S}_{2,u+t_y} + J_x~\vec{S}_{2,u} \cdot \vec{S}_{4,u-t_x-t_y} \nonumber\\
&+& J_y~\vec{S}_{1,u} \cdot \vec{S}_{4,u-t_x-t_y} + J_y~\vec{S}_{4,u} \cdot \vec{S}_{1,u+t_x}
\eea
where $\vec{t}_{x,y} = 2 a~ \vec{e}_{x,y}$ are the primitive translation vectors in the model ($a$ is the lattice constant of the square lattice), and  the index $u$ labels the unit cell.
The remaining spin operator index (1-4) labels the 4 sites of the unit cell (see right panel of Fig.~\ref{fig:PertTheory}).
We then take as our reference classical state the one where the spins 1 and 3 (resp. 2 and 4) of the unit cell 
point along the positive (resp. negative) $z$-axis, and perform a standard Holstein-Primakoff expansion
in terms of four bosonic operators labeled as $a_u, b_u, c_u, d_u$, which correspond to the four spins per unit cell $S_{1,u}, S_{2,u}, S_{3,u}, S_{4,u}$ respectively. 
The transformation reads
\bea
&& S_{1,u}^z \simeq  -S + a_u^+ a_u, ~~~ S_{1,u}^+ \simeq \sqrt{2 S} ~a_u^+ \nonumber \\
&& S_{2,u}^z \simeq  S - b_u^+ b_u, ~~~ S_{2,u}^+ \simeq \sqrt{2 S} ~b_u \nonumber \\
&& S_{3,u}^z \simeq  S - c_u^+ c_u, ~~~ S_{3,u}^+ \simeq \sqrt{2 S} ~c_u \nonumber \\
&& S_{4,u}^z \simeq  -S + d_u^+ d_u, ~~~ S_{4,u}^+ \simeq \sqrt{2 S} ~d_u^+ ~.
\eea
Replacing these expressions in the Hamiltonian and performing a Fourier
transform, we get 
\be
\mc{H} = E_{0b} +  \mc{H}_{2b} + \mc{O} (S^0)
\ee
where $E_{ob} = -2 \xi N_{uc} S^2$ stands for the classical energy, and $\xi \equiv J + 2 J_x - 2 J_y$.  
The term $\mc{H}_{2b}$ stands for the quadratic Hamiltonian which can be written in the compact matrix form  
\be
\mc{H}_{2b} = \delta E_1 + \frac{S}{2} \sum_k \vec{A}_k^+ \cdot \vec{M}_k \cdot \vec{A}_k
\ee
where $\delta E_1 =  -2 \xi N_{uc} S$,  
\be
\vec{A}_k^+ \!=\! \left(
\begin{array}{cccccccc} 
a_k^+, & b_k^+, & c_k^+, & d_k^+, & a_{-k}, & b_{-k}, & c_{-k}, & d_{-k} 
\end{array} \right), 
\ee
\be
\vec{M}_k = \left(\begin{array}{cc}
\vec{C}_k & \vec{D}_k \\
\vec{D}_k & \vec{C}_k
\end{array}\right) ~,
\ee 
and
\begin{widetext} 
{\small 
\be
\vec{C}_k = \left(
\begin{array}{cccc}
\xi & 0 & 0 & J_y e^{-i k_x} (1+e^{-i k_y}) \\ 
&&&\\
0 & \xi & J_y (1+e^{-i k_y}) & 0\\ 
&&&\\
0 &  J_y (1+e^{i k_y})& \xi & 0  \\ 
&&&\\
 J_y e^{i k_x}(1+e^{i k_y}) &  0 & 0 & \xi 
\end{array}
\right)~,
\ee
\be
\vec{D}_k = \left(
\begin{array}{cccc}
0 &  -J & -J_x (1+e^{-i k_x}) & 0\\
&&&\\
-J & 0 & 0 & -J_x e^{-i k_y}(1+e^{-i k_x})\\
&&&\\
 -J_x (1+e^{i k_x}) & 0 & 0 & -J\\
&&&\\
0 & -J_x e^{i k_y}(1+e^{i k_x}) & -J  & 0\\
\end{array} \right)~. 
\ee}
\end{widetext}
To diagonalize $\mc{H}_{2b}$ we search for a new set of bosonic operators $\tilde{\vec{A}}_k$ 
given by the generalized Bogoliubov transformation $\vec{A}_k = \vec{S}_k \cdot \tilde{\vec{A}}_k$, such that   
the matrix $\vec{S}_k^+  \vec{M}_k \vec{S}_k \equiv \vec{\Omega}_k$ becomes diagonal. 
The transformation must also preserve the bosonic commutation relations, which can be expressed compactly as 
$\vec{g}=\tilde{\vec{g}} = \vec{S}_k \cdot \vec{g}\cdot  \vec{S}_k^+ $, where $\vec{g}$ is the ``commutator'' matrix 
\be
\vec{g} = \vec{A}_k \cdot \vec{A}_k^+ - \left( \left(\vec{A}_k^+\right)^T \cdot \vec{A}_k^T \right)^T= 
\left(\begin{array}{cc}
{\bf 1}_4 & 0 \\
0 & -{\bf 1}_4 
\end{array}\right) 
\ee
and ${\bf 1}_4$ stands for the $4\times 4$ identity matrix. The above two conditions give 
\be
(\vec{g} \vec{M}_k)\cdot \vec{S}_k = \vec{S}_k \cdot \left( \vec{g} \vec{\Omega}_k \right) \equiv \vec{S}_k \cdot \vec{\Omega}'_k  
\ee
which is an eigenvalue equation in matrix form (the columns of $\vec{S}_k$ contain the eigenvectors of $\vec{g} \vec{M}_k$).  

One can show\cite{Blaizot} that if $\vec{M}_k$ is semi-definite positive, then the eigenvalues of $\vec{g M}_k$ come in opposite pairs: 
\be
\vec{\Omega}'_k = \left(\begin{array}{cc}
\boldsymbol{\omega}_k & 0 \\
0 & -\boldsymbol{\omega}_k 
\end{array}\right)
\Rightarrow 
\vec{\Omega}_k = \left(\begin{array}{cc}
\boldsymbol{\omega}_k & 0 \\
0 & \boldsymbol{\omega}_k 
\end{array}\right)  
\ee
where $\boldsymbol{\omega}_k$ is a diagonal matrix with non-negative entries $\omega_{1k}, \!\ldots\!, \omega_{4k}$. This in turn leads to 
\be
\mc{H}_{2b} =\delta E_1 + \delta E_2 + S \sum_{k} \left( \omega_{1k}~ \tilde{a}_{k}^+ \tilde{a}_{k} +\! \ldots\! +\omega_{4k}~ \tilde{d}_{k}^+ \tilde{d}_{k} \right) ~,
\ee   
where the second correction 
\be\label{eqn:dE2} 
\delta E_2 = \frac{S}{2} \sum_{k}\left( \omega_{1k}+\!\ldots\!+ \omega_{4k}\right)~,
\ee 
stands for the total contribution from the zero-point energy of all independent harmonic operators involved in the theory. 

\subsection{Renormalization of the magnetic moments}
Let us now look at the effect of the quadratic fluctuations on the
magnetic moments.  We consider the spin $\vec{S}_1$ operator inside the
unit cell $n=0$.  The classical vector points along the $z$-axis. We
have
\be
\vec{S}_1^z= S-a_{n=0}^+ a_{n=0} = S-\frac{1}{N_{uc}} \sum_{k,q} a_k^+ a_q   ~.
\ee
Using the transformation $\vec{A}_k = \vec{S}_k\cdot\tilde{\vec{A}}_k$, we get
\be
\vec{S}_1^z \!=\! S-\frac{1}{N_{uc}} \sum_{k,q} \sum_{ij} S_k(1,i)^\ast S_q(1,j) ~\tilde{A}_k^+(i) \tilde{A}_k(j)~.
\ee
In the vacuum GS, the only non-vanishing expectation values are of the type $\langle \tilde{a}_k \tilde{a}_k^+ \rangle=1$.
Thus from the above sum we should keep only terms with $i\!=\!j\!=\!5-8$ and $k\!=\!q$, namely
\be\label{eqn:dSz}
\langle \vec{S}_1^z \rangle = S-\frac{1}{N_{uc}} \sum_k \sum_{i=5}^8 |S_k(1,i)|^2~.
\ee
The second term can be calculated by a numerical integration over the BZ. 

%

\end{document}